\documentclass[journal,draftcls,onecolumn,12pt,twoside]{IEEEtran}

\ifCLASSINFOpdf
\usepackage[pdftex]{graphicx}
  \graphicspath{{../pdf/}{../jpeg/}}
  \DeclareGraphicsExtensions{.pdf,.jpeg,.png}
\else
  \usepackage[dvips]{graphicx}
  \graphicspath{{../eps/}}
  \DeclareGraphicsExtensions{.eps}
\fi
\usepackage{epstopdf}
\usepackage{amsmath,amsthm, amssymb}
\usepackage{mathrsfs}
\usepackage{threeparttable}
\usepackage{bm}
\usepackage{multirow}
\usepackage{booktabs}
\usepackage{color}
\usepackage{tabularx}
\usepackage{setspace}
\usepackage{verbatim}
\usepackage{stfloats}
\usepackage{caption}
\usepackage{float}
\usepackage[T1]{fontenc}
\usepackage{soul}
\usepackage{cite}

\usepackage{stfloats}
\usepackage{mathtools}
\usepackage{amsmath}
\allowdisplaybreaks[0]
\usepackage{subfigure}
\usepackage{fancyhdr}

\makeatletter
\newif\if@restonecol
\makeatother

\usepackage[linesnumbered,ruled,vlined]{algorithm2e}%[ruled,vlined]{
\usepackage{algpseudocode}
\usepackage{amsmath}
\usepackage{color}

\begin{document}

\title{Mobility-Aware Cooperative Caching in Vehicular Edge Computing Based on Asynchronous Federated and Deep Reinforcement Learning}

\author{Qiong Wu,~\IEEEmembership{Member,~IEEE}, Yu Zhao, Qiang Fan, Pingyi Fan, ~\IEEEmembership{Senior Member,~IEEE},\\Jiangzhou Wang,~\IEEEmembership{Fellow,~IEEE}, Cui Zhang
\thanks{This work was supported in part by the National Natural Science Foundation of China under Grant No. 61701197, in part by the open research fund of State Key Laboratory of Integrated Services Networks under Grant No. ISN23-11,  in part by the 111 Project under Grant No. B12018. \emph{(Corresponding author: Qiong Wu)}

Qiong Wu and Yu Zhao are with School of Internet of Things Engineering, Jiangnan University, Wuxi 214122, China, and also with the State Key Laboratory of Integrated Services Networks (Xidian University),  Xi'an 710071, China (Email: qiongwu@jiangnan.edu.cn, yuzhao@stu.jiangnan.edu.cn).

Qiang Fan is with Qualcomm, San Jose CA 95110 USA (Email: qiangfan29@gmail.com).

Pingyi Fan is with the Department of Electronic Engineering, Beijing National Research Center for Information Science and Technology, Tsinghua University, Beijing 100084, China (Email: fpy@tsinghua.edu.cn).

Jiangzhou Wang is with the School of Engineering, University of Kent, CT2 7NT Canterbury, U.K. (Email: j.z.wang@kent.ac.uk).

Cui Zhang is with Banma Network Technology Co., Ltd., Shanghai 200000, China (Email: zc351340@alibaba-inc.com).

}
}
% <-this % stops a space
% \thanks{Manuscript received XXX, XX, 20; revised XXX, XX, 2015.}}

\markboth{}
{}
% {Shell \MakeLowercase{\textit{et al.}}: Bare Demo of IEEEtran.cls for Journals}

\maketitle

\begin{abstract}
The vehicular edge computing (VEC) can cache contents in different RSUs at the network edge to support the real-time vehicular applications. In VEC, owing to the high-mobility characteristics of vehicles, it is necessary to cache the user data in advance and learn the most popular and interesting contents for vehicular users. Since user data usually contains privacy information, users are reluctant to share their data with others. To solve this problem, traditional federated learning (FL) needs to update the global model synchronously through aggregating all users' local models to protect users' privacy. However, vehicles may frequently drive out of the coverage area of the VEC before they achieve their local model trainings and thus the local models cannot be uploaded as expected, which would reduce the accuracy of the global model. In addition, the caching capacity of the local RSU is limited and the popular contents are diverse, thus the size of the predicted popular contents usually exceeds the cache capacity of the local RSU. Hence, the VEC should cache the predicted popular contents in different RSUs while considering the content transmission delay. In this paper, we consider the mobility of vehicles and propose a cooperative Caching scheme in the VEC based on Asynchronous Federated and deep Reinforcement learning (CAFR). We first consider the mobility of vehicles and propose an asynchronous FL algorithm to obtain an accurate global model, and then propose an algorithm to predict the popular contents based on the global model. In addition, we consider the mobility of vehicles and propose a deep reinforcement learning algorithm to obtain the optimal cooperative caching location for the predicted popular contents in order to optimize the content transmission delay. Extensive experimental results have demonstrated that the CAFR scheme outperforms other baseline caching schemes.

%In the VEC, the local RSU should predict the popular contents in the high-mibility vehicular environment to reduce the content transmission delay. Federated learning (FL) enables the local RSU to aggregate vehicles' local models instead of data to update the global model, thus vehicles can avoid from sharing data. Traditional synchronous FL algorithm needs to wait for all vehicles to finish training and upload their local models before aggregation. However, some vehicles may drive out of the coverage of the local RSU before they finish training and thus cannot upload their local models, which would reduce the accuracy of the global model, and further prolong the content transmission delay. Furthermore, the local RSU and neighbour RSU cache different popular contents would cause different content transmission delay. In this paper, we consider the mobility of vehicles and propose a cooperative caching scheme in the VEC based on asynchronous federated and deep reinforcement learning (CAFR). We first propose an asynchronous FL algorithm to obtain an accurate global model, and then propose an algorithm to predict the popular contents based on the global model. In addition, we propose a caching scheme to minimizes the content transmission delay based on the dueling deep Q-network (DQN) algorithm. Experimental results show that the CAFR scheme outperforms other baseline caching schemes in terms of cache hit rate and average content transmission delay.

\end{abstract}

\IEEEpeerreviewmaketitle
\section{Introduction}
\label{sec1}
\IEEEPARstart{W}{ith} the development of the internet of vehicles (IoV) and cloud computing, caching technology facilitates various real-time vehicular applications for vehicular users (VUs), such as automatic navigation, pattern recognition and multimedia entertainment \cite{Liuchen2021} \cite{QWu2022}. For the standard caching technology, the cloud caches various contents like data, video and web pages. In this scheme, vehicles transmit the required contents to a macro base station (MBS) connected to a cloud server, and could fetch the contents from the MBS, which would cause high content transmission delay from the MBS to vehicles due to the communication congestion caused by frequently requested contents from vehicles \cite{Dai2019}. The content transmission delay can be effectively reduced by the emergence of vehicular edge computing (VEC), which caches contents in the road side unit (RSU) deployed at the edge of vehicular networks (VNs) \cite{Javed2021}. Thus, vehicles can fetch contents directly from the local RSU, to reduce the content transmission delay. In the VEC, since the caching capacity of the local RSU is limited, if some vehicles cannot fetch their required contents, a neighboring RSU who has the required contents could forward them to the local RSU. The worst case is that vehicles need to fetch contents from the MBS due to both local and neighboring RSUs not having cached the requested contents.

In the VEC, it is critical to design a caching scheme to cache the popular contents. The traditional caching schemes cache contents based on the previously requested contents \cite{Narayanan2018}. However, owing to the high-mobility characteristics of vehicles in VEC, the previously requested contents from vehicles may become outdated quickly, thus the traditional caching schemes may not satisfy all the VUs' requirement. Therefore, it is necessary to predict the most popular contents in the VEC and cache them in the suitable RSUs in advance. Machine learning (ML) as a new tool, can extract hidden features by training user data to efficiently predict popular contents\cite{Yan2019}. However, the user data usually contains privacy information and users are reluctant to share their data directly with others, which make it difficult to collect and train users' data. Federated learning (FL) can protect the privacy of users by sharing their local models instead of data\cite{Chen2021}. In traditional FL, the global model is periodically updated by aggregating all vehicles' local models\cite{Wang2020} -\cite{Cheng2021}. However, vehicles may frequently drive out of the coverage area of the VEC before they update their local models and thus the local models cannot be uploaded in the same area, which would reduce the accuracy of the global model as well as the probability of getting the predicted popular contents. Hence, it motivates us to consider the mobility of vehicles and propose an asynchronous FL to predict accurate popular contents in VEC.

Generally, the predicted popular contents should be cached in their local RSU of vehicles to guarantee a low content transmission delay. However, the caching capacity of each local RSU is limited and the popular contents may be diverse, thus the size of the predicted popular contents usually exceeds the cache capacity of the local RSU. Hence, the VEC has to determine where the predicted popular contents are cached and updated. The content transmission delay is an important metric for vehicles to provide real-time vehicular application. The different popular contents cached in the local and neighboring RSUs would impact the way vehicles fetch contents, and thus affect the content transmission delay. In addition, the content transmission delay of each vehicle is impacted by its channel condition, which is affected by vehicle mobility. Therefore, it is necessary to consider the mobility of vehicles to design a cooperative caching scheme, in which the predicted popular contents can be cached among RSUs to optimize the content transmission delay. In contrast to some conventional decision algorithms, deep reinforcement learning (DRL) is a favorable tool to construct the decision-making framework and optimize the cooperative caching for the contents in complex vehicular environment \cite{Zhu2021}. Therefore, we shall employ DRL to determine the optimal cooperative caching to reduce the content transmission delay of vehicles.

In this paper, we consider the vehicle mobility and propose a cooperative Caching scheme in VEC based on Asynchronous Federated and deep Reinforcement learning (CAFR). The main contributions of this paper are summarized as follows.

\begin{itemize}
\item[1)] By considering the mobility characteristics of vehicles including the positions and velocities, we propose an asynchronous FL algorithm to improve the accuracy of the global model.

%asynchronous FL allows vehicles' data to be trained locally thereby reducing the privacy risk of vehicles, lowering communication costs, and adapting to highly dynamic VN environments. After training, the hidden features of users and contents can be obtained which are used to predicted its content popularity.
\item[2)] We propose an algorithm to predict the popular contents based on the global model, where each vehicle adopts the autoencoder (AE) to predict its interested contents based on the global model, while the local RSU collects the interested contents of all vehicles within the coverage area to catch the popular contents.

\item[3)] We elaborately design a DRL framework (dueling deep Q-network (DQN)) to illustrate the cooperative caching problem, where the state, action and reward function have been defined. Then the local RSU can determine optimal cooperative caching to minimize the content transmission delay based on the dueling DQN algorithm.
\end{itemize}

The rest of the paper is organized as follows. Section \ref{sec2} reviews the related works on content caching in VNs. Section \ref{sec3} briefly describes the system model. Section \ref{sec5} proposes a mobility-aware cooperative caching in the VEC based on asynchronous federated and deep reinforcement learning method. We present some simulation results in Section \ref{sec6}, and then conclude them in Section \ref{sec7}.

\section{Related Work}
\label{sec2}
In this section, we first review the existing works related to the content caching in vehicular networks (VNs), and then survey the current state of art of the cooperative content caching schemes in VEC.

In \cite{YDai2020}, Dai \textit{et al.} proposed a distributed content caching framework with empowering blockchain to achieve security and protect privacy, and considered the mobility of vehicles to design an intelligent content caching scheme based on DRL framework.
In \cite{Yu2021}, Yu \textit{et al.} proposed a mobility-aware proactive edge caching scheme in VNs that allows multiple vehicles with private data to collaboratively train a global model for predicting content popularity, in order to meet the requirements for computationally intensive and latency-sensitive vehicular applications.
In \cite{JZhao2021}, Zhao \textit{et al.} optimized the edge caching and computation management for service caching, and adopted Lyapunov optimization to deal with the dynamic and unpredictable challenges in VNs.
In \cite{SJiang2020}, Jiang \textit{et al.} constructed a two-tier secure access control structure for providing content caching in VNs with the assistance of edge devices, and proposed the group signature-based scheme for the purpose of anonymous authentication.
In \cite{CTang2021}, Tang \textit{et al.} proposed a new optimization method to reduce the average response time of caching in VNs, and then adopted Lyapunov optimization technology to constrain the long-term energy consumption to guarantee the stability of response time.
In \cite{YDai2022}, Dai \textit{et al.} proposed a VN with digital twin to cache contents for adaptive network management and policy arrangement, and designed an offloading scheme based on the DRL framework to minimize the total offloading delay.
However, the above content caching schemes in VNs did not take into account the cooperative caching in the VEC environment.

There are some works considering cooperative content caching schemes in VEC.
In \cite{GQiao2020}, Qiao \textit{et al.} proposed a cooperative edge caching scheme in VEC and constructed the double time-scale markov decision process to minimize the content access cost, and employed the deep deterministic policy gradient (DDPG) method to solve the long-term mixed-integer linear programming problems.
In \cite{JChen2020}, Chen \textit{et al.} proposed a cooperative edge caching scheme in VEC which considered the location-based contents and the popular contents, while designing an optimal scheme for cooperative content placement based on an ant colony algorithm to minimize the total transmission delay and cost.
In \cite{LYao2022}, Yao \textit{et al.} designed a cooperative edge caching scheme with consistent hash and mobility prediction in VEC to predict the path of each vehicle, and also proposed a cache replacement policy based on the content popularity to decide the priorities of collaborative contents.
In \cite{RWang2021}, Wang \textit{et al.} proposed a cooperative edge caching scheme in VEC based on the long short-term memory (LSTM) networks, which caches the predicted contents in RSUs or other vehicles and thus reduces the content transmission delay.
In \cite{DGupta2020}, Gupta \textit{et al.} proposed a cooperative caching scheme that jointly considers cache location, content popularity and predicted rating of contents to make caching decision based on the non-negative matrix factorization, where it employs a legitimate user authorization to ensure the secure delivery of cached contents.
In \cite{LYao2019}, Yao \textit{et al.} proposed a cooperative caching scheme based on the mobility prediction and drivers' social similarities in VEC, where the regularity of vehicles' movement behaviors are predicted based on the hidden markov model to improve the caching performance.
In \cite{RWu2022}, Wu \textit{et al.} proposed a hybrid service provisioning framework and cooperative caching scheme in VEC to solve the profit allocation problem among the content providers (CPs), and proposed an optimization model to improve the caching performance in managing the caching resources.
In \cite{LYao2017}, Yao \textit{et al.} proposed a cooperative caching scheme based on mobility prediction, where the popular contents may be cached in the mobile vehicles within the coverage area of hot spot. They also designed a cache replacement scheme according to the content popularity to solve the limited caching capacity problem for each edge cache device.
In \cite{KZhang2018}, Zhang \textit{et al.} proposed a cooperative edge caching architecture that focuses on the mobility-aware caching, where the vehicles cache the contents with base stations collaboratively. They also introduced a vehicle-aided edge caching scheme to improve the capability of edge caching.
In \cite{KLiu2016}, Liu \textit{et al.} designed a cooperative caching scheme that allows vehicles to search the unrequested contents. This scheme facilitates the content sharing among vehicles and improves the service performance.
In \cite{SWang2017}, Wang \textit{et al.} proposed a VEC caching scheme to reduce the total transmission delay. This scheme extends the capability of the data center from the core network to the edge nodes by cooperatively caching popular contents in different CPs. It minimizes the VUs' average delay according to an iterative ascending price method.
In \cite{MLiu2021}, Liu \textit{et al.} proposed a real-time caching scheme in which edge devices cooperate to improve the caching resource utilization. In addition, they adopted the DRL framework to optimize the problem of searching requests and utility models to guarantee the search efficiency.
In \cite{BKo2019}, Ko \textit{et al.} proposed an adaptive scheduling scheme consisting of the centralized scheduling mechanism, ad hoc scheduling mechanism and cluster management mechanism to exploit the ad hoc data sharing among different RSUs.
In \cite{JCui2020}, Cui \textit{et al.} proposed a privacy-preserving data downloading method in VEC, where the RSUs can find popular contents by analyzing encrypted requests of nearby vehicles to improve the downloading efficiency of the network.
In \cite{QLuo2020}, Luo \textit{et al.} designed a communication, computation and cooperative caching framework, where computing-enabled RSUs provide computation and bandwidth resource to the VUs to minimize the data processing cost in VEC.

As mentioned above, no other works has considered the vehicle mobility and privacy of VUs simultaneously to design cooperative caching schemes in VEC, which motivates us to propose a mobility-aware cooperative caching in VEC based on the asynchronous FL and DRL.

\begin{figure}
\center
\includegraphics[scale=0.7]{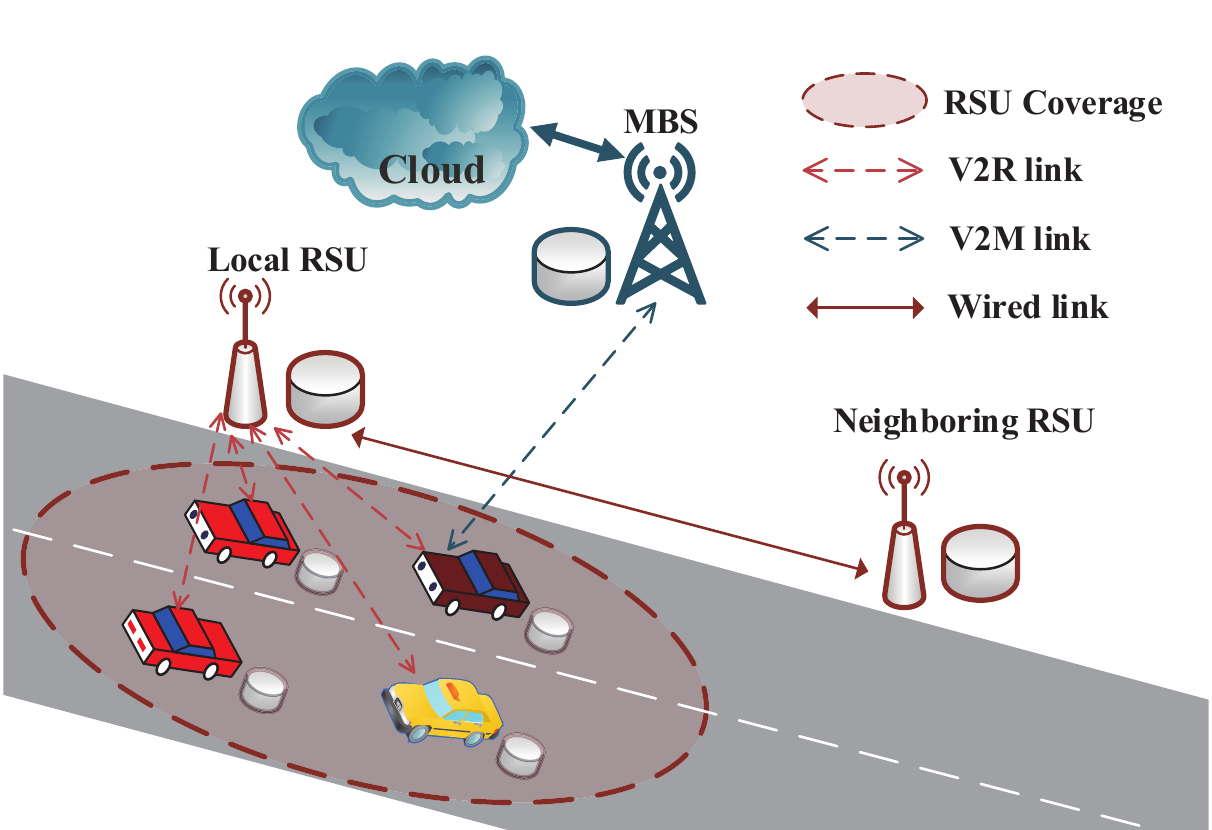}
\caption{VEC scenario}
\label{fig1}
\end{figure}

\section{System Model}
\label{sec3}
%In this section, we first introduce the system scenario, and then describe the mobility model of VUs. Finally, we introduce the communication model. The notations in this paper are summarized in Table \ref{tab1}.

\subsection{System Scenario}
As shown in Fig. \ref{fig1}, we consider a three-tier VEC in an urban scenario that consists of a local RSU, a neighboring RSU, a MBS attached with a cloud and some vehicles moving in the coverage area of the local RSU. The top tier is the MBS deployed at the center of the VEC, while middle tier is the RSUs deployed in the coverage area of the MBS. They are placed on one side of the road. The bottom tier is the vehicles driving within the coverage area of the RSUs.
%The RSUs are denoted as $\{S_1,S_2,...,S_i,...,S_N\}$, where $S_i$ is the $i$th RSU $(1 \leq i \leq N)$.

%The coverage range of the local RSU is $L_s$.
%Different vehicles carry different number of users.
%the rating for a content is a value within [0,1].

Each vehicle stores a large amount of VUs' historical data, i.e., local data. Each data is a vector reflecting different information of a VU, including the VU's personal information such as identity (ID) number, gender, age and postcode, the contents that the VU may request, as well as the VU's ratings for the contents where a larger rating for a content indicates that the VU is more interested in the content. Particularly, the rating for a content may be $0$, which means that it is not popular or is not requested by VUs. Each vehicle randomly chooses a part of the local data to form a training set while the rest is used as a testing set. The time duration of vehicles within the coverage area of the MBS is divided into rounds. For each round, each vehicle randomly selects contents from its testing set as the requested contents, and sends the request information to the local RSU to fetch the contents at the beginning of each round. We consider the MBS has abundant storage capacity and caches all available contents, while the limited storage capacity of each RSU can only accommodate part of contents. Therefore, the vehicle fetches each of the requested content from the local RSU, neighboring RSU or MBS in different conditions. Specifically,

\subsubsection{Local RSU}If a requested content is cached in the local RSU, the local RSU sends back the requested content to the vehicle. In this case the vehicle fetches the content from the local RSU.
\subsubsection{neighboring RSU}If a requested content is not cached in the local RSU, the local RSU transfers the request to the neighboring RSU, and the neighboring RSU sends the content to the local RSU if it caches the requested content. Afterward, the local RSU sends back the content to the vehicle. In this case the vehicle fetches the content from the neighboring RSU.
\subsubsection{MBS}If a content is neither cached in the local RSU nor the neighboring RSU, the vehicle sends the request to the MBS that directly sends back the requested content to the vehicle. In this case, the VU fetches the content from the MBS.

\subsection{Mobility Model of Vehicles}
The model assumes that all vehicles drive in the same direction and vehicles arrive at a local RSU, following a Poisson distribution with the arrival rate $\lambda_{v}$. Once a vehicle enters the coverage of the local RSU, it sends request information to the local RSU. Each vehicle keeps the same mobility characteristics including position and velocity within a round and may change its mobility characteristics at the beginning of each round. The velocity of different vehicles follows an independent identically distribution. The velocity of each vehicle is generated by a truncated Gaussian distribution, which is flexible and consistent with the real dynamic vehicular environment. For round $r$, the number of vehicles driving in the coverage area of the local RSU is $N^{r}$. The set of $N^{r}$ vehicles are denoted as $\mathbb{V}^{r}=\left\{V_{1}^{r}, V_{2}^{r},\ldots, V_{i}^{r}, \ldots, V_{N^{r}}^{r}\right\}$, where $V_{i}^{r}$ is vehicle $i$ driving in the local RSU $(1 \leq i \leq N^{r})$. Let $\left\{U_{1}^{r}, U_{2}^{r}, \ldots, U_{i}^{r}, \ldots, U_{N^{r}}^{r}\right\}$ be the velocities of all vehicles driving in the local RSU, where $U_{i}^{r}$ is velocity of $V_{i}^{r}$. According to \cite{AlNagar2019}, the probability density function of $U_{i}^{r}$ is expressed as
\begin{equation}
f({U_{i}^r}) = \left\{ \begin{aligned}
\frac{{{e^{ - \frac{1}{{2{\sigma ^2}}}{{({U_{i}^r} - \mu )}^2}}}}}{{\sqrt {2\pi {\sigma ^2}} (erf(\frac{{{U_{\max }} - \mu }}{{\sigma \sqrt 2 }}) - erf(\frac{{{U_{\min }} - \mu }}{{\sigma \sqrt 2 }}))}},\\
{U_{min }} \le {U_{i}^r} \le {U_{max }},\\
0 \qquad \qquad \qquad \qquad \quad otherwise.
\end{aligned} \right.
\label{eq1}
\end{equation}
where $U_{\max}$ and $U_{\min}$ are the maximum and minimum velocity threshold of each vehicle, respectively, and $erf\left(\frac{U_{i}^{r}-\mu}{\sigma \sqrt{2}}\right)$ is the Gauss error function of $U_{i}^{r}$ under the mean $\mu$ and variance $\sigma^{2}$.
\subsection{Communication Model}

The communications between the local RSU and neighboring RSU adopt the wired link. Each vehicle keeps the same communication model during a round and changes its communication model for different rounds. When the round is $r$, the channel gain of $V_{i}^{r}$ is modeled as \cite{3gpp}

\begin{equation}
\begin{aligned}
h_{i}^{r}(dis(x,V_{i}^{r}))=\alpha_{i}^{r}(dis(x,V_{i}^{r})) g_{i}^{r}(dis(x,V_{i}^{r})), \\
x=S,M,\\
\label{eq2}
\end{aligned}
\end{equation}
where $x=S$ means the local RSU and $x=M$ means the MBS, $dis(x,V_{i}^{r})$ is the distance between the local RSU$/$MBS and $V_{i}^{r}$, $\alpha_{i}^{r}(dis(x,V_{i}^{r}))$ is the path loss between the local RSU$/$MBS and $V_{i}^{r}$, and $g_{i}^{r}(dis(x,V_{i}^{r}))$ is the shadowing channel fading between the local RSU$/$MBS and $V_{i}^{r}$, which follows a Log-normal distribution.

Each RSU communicates with the vehicles in its coverage area through vehicle to RSU (V2R) link, while the MBS communicates with vehicles through vehicle to base station (V2B) link. Since the distances between the local RSU$/$MBS and $V_{i}^{r}$ are different in different rounds, V2R$/$V2B link suffers from different channel impairments, and thus transmit with different transmission rates in different rounds. The transmission rates under V2R and V2B link are calculated as follows.

According to the Shannon theorem, the transmission rate between the local RSU and $V_{i}^{r}$ is calculated as \cite{Chenwu2020}
\begin{equation}
R_{R, i}^{r}=B\log _{2}\left(1+\frac{p_B h_{i}^{r}(dis(S,V_{i}^{r}))}{\sigma_{c}^{2}}\right),
\label{eq3}
\end{equation}where $B$ is the available bandwidth, $p_B$ is the transmit power level used by the local RSU and $\sigma_{c}^{2}$ is the noise power.

Similarly, the transmission rate between the MBS and $V_{i}^{r}$ is calculated as

\begin{equation}
R_{B, i}^{r}=B\log _{2}\left(1+\frac{p_{M} h_{i}^{r}(dis(M,V_{i}^{r}))}{\sigma_{c}^{2}}\right),
\label{eq4}
\end{equation}where $p_{M}$ is the transmit power level used by MBS.

%In the cooperative caching network we have discussed,

\begin{figure}
\center
\includegraphics[scale=0.75]{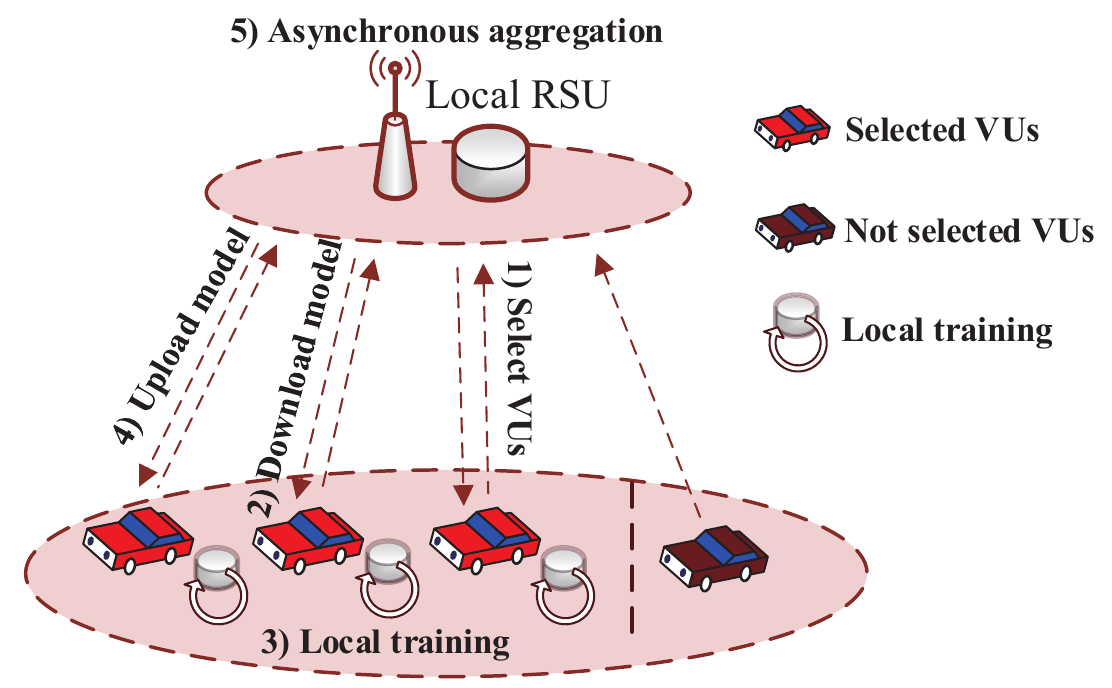}
\caption{Asynchronous FL}
\label{fig2}
\end{figure}

\section{Cooperative Caching Scheme}
\label{sec5}
In this section, we propose a cooperative cache scheme to optimize the content transmission delay in each round $r$. We first propose an asynchronous FL algorithm to protect VU's information and obtain an accurate model. Then we propose an algorithm to predict the popular contents based on the obtained model. Finally, we present a DRL based algorithm to determine the optimal cooperative caching according to the predicted popular contents. Next, we will introduce the asynchronous FL algorithm, the popular content prediction algorithm and the DRL-based algorithm, respectively.

\subsection{Asynchronous Federated Learning}
%describe the cooperative caching scheme based on MCFR in detail. We first introduce asynchronous federated learning framework which includes VU selection, global model download, local model training, updated model upload and asynchronous aggeration method. Then, we introduce the process about predicting the content popularity. Finally, with the predicted content popularity and the content transmission delay of all VUs, we adopt the dueling deep Q-Network (DQN) to explore a cooperative content placement scheme.

%FL promotes collaborative training of deep neural network model among VUs within the coverage area of the RSU by storing data on the VUs, which significantly reduces the privacy risk of VUs and the communication costs. Most works have adopted the synchronous FL framework, which the RSU needs to wait for all VUs to send its updated model before aggregation, the stragglers will lead to the failure to get an accurate global model in highly dynamic VN scenario. Therefore we adopt the asynchronous FL framework, in contrast to synchronous FL, the RSU updates the global model by weighted averaging after receiving updates from a VU without waiting for other VUs to finish training local models.
As shown in Fig. \ref{fig2}, the asynchronous FL algorithm consists of 5 steps as follows.

%Specifically, 1) Select VUs: RSU $S_i$ first selects VUs which has enough time to participate the asynchronous FL according to the VUs' mobility characteristics including the velocity and position. 2) Download model: each selected VU download the global model from $S_i$. 3) Local model training: each selected VU adopts its local data to train the local model. 4) Upload model: $S_i$ receives a local model of a VU with the fastest training time among all selected VUs. 5) Asynchronous aggregation: $S_i$ aggregates the updated local model and the previous global model to update the global model. The designed asynchronous FL algorithm is shown in Fig. \ref{fig2} and the 5 steps of the asynchronous FL algorithm are specified as follows:

\subsubsection{Select Vehicles}
\
\newline
\indent
%Since the coverage area of RSU is limited and VUs on the urban road have a certain speed, there may be a few VUs that cannot complete asynchronous FL training process due to the short staying time when they cross the current local RSU, and this situation will result in the inefficient global model trained by asynchronous FL in each RSU, which will make the inefficient caching performance. Updating the aggregation of the high-quality VU's models in each RSU server can train a more accurate global model, and the VU that passes the screening is used as a node to compute the local data for updating the global model of local RSU.
The main goal of this step is to select the vehicles whose staying time in the local RSU is long enough to ensure they can participate in the asynchronous FL and complete the training process.

Each vehicle first sends its mobility characteristics including its velocity and position (i.e., the distance to the local RSU and distance it has traversed within the coverage of the local RSU), then the local RSU selects vehicles according to the staying time that is calculated based on the vehicle's mobility characteristics. The staying time of $V_{i}^{r}$ in the local RSU is calculated as

\begin{equation}
T_{r,i}^{staying}=\left(L_{s}-P_{i}^{r}\right) / U_{i}^{r},
\label{eq5}
\end{equation}
where $L_s$ is the coverage range of the local RSU, $P_{i}^{r}$ is the distance that $V_{i}^{r}$ has traversed within the coverage of the local RSU.

The staying time of $V_{i}^{r}$ should be larger than the sum of the average training time $T_{training}$ and inference time $T_{inference}$ to guarantee that  $V_{i}^{r}$ can complete the training process. Therefore, if $T_{r,i}^{staying}>T_{training}+T_{inference}$, the local RSU selects $V_{i}^{r}$ to participate in asynchronous FL training. Otherwise, $V_{i}^{r}$ is ignored.

\subsubsection{Download Model}
\
\newline
\indent
In this step, the local RSU will generate the global model $\omega^{r}$. For the first round,  the local RSU initializes a global model based on the AE, which can extract the hidden features used for popular content prediction. In each round, the local RSU updates the global model and transfers the global model $\omega^{r}$ to all the selected vehicles in the end.

%of the selected VUs participating in asynchronous FL training in the previous round, and updates the global model based on it. Employing the previous model improves the efficiency of model training and saves training time.

% Define $N_{r}$ as the total number of vehicles selected for asynchronous FL training when the communication round is $r$.

\subsubsection{Local Training}
\
\newline
\indent
In this step, each vehicle in the local RSU sets the downloaded global model $\omega^{r}$ as the initial local model and updates the local model iteratively through training. Afterward, the updated local model will be the feedback to the local RSU.
For each iteration $k$, $V_{i}^{r}$ randomly samples some training data $n_{i,k}^{r}$ from the training set. Then, it uses $n_{i,k}^{r}$ to train the local model based on the AE that consists of an encoder and a decoder. Let $W_{i,k}^{r,e}$ and $b_{i,k}^{r,e}$ be the weight matrix and bias vector of the encoder for iteration $k$, respectively, $W_{i,k}^{r,d}$ and $b_{i,k}^{r,d}$ be the weight matrix and bias vector of the decoder for iteration $k$, respectively. Thus the local model of $V_{i,j}^{r}$ for iteration $k$ is expressed as $\omega_{i,k}^r=\{W_{i,k}^{r,e}, b_{i,k}^{r,e}, W_{i,k}^{r,d}, b_{i,k}^{r,d}\}$. For each training data $x$ in $n_{i,k}^{r}$, the encoder first maps the original training data $x$ to a hidden layer to obtain the hidden feature of $x$, i.e., $z(x)=f\left(W_{i,k}^{r,e}x+b_{i,k}^{r,e}\right)$. Then the decoder calculates the reconstructed input $\hat{x}$, i.e., $\hat{x}=g\left(W_{i,k}^{r,d}z(x)+b_{i,k}^{r,d}\right)$, where $f{(\cdot)}$ and $g{(\cdot)}$ are the nonlinear and logical activation function \cite{Ng2011}. Afterward, the loss function of data $x$ under the local model $\omega_{i,k}^r$ is calculated as
\begin{equation}
l\left(\omega_{i,k}^r;x\right)=(x-\hat{x})^{2},
\label{eq6}
\end{equation}where $\omega^{r}_{i,1}=\omega^{r}$.

%If the reconstruction loss is lower, it means that our model is better at extracting features between data and reconstructing the data well enough to predict the unknown data.

%, which is the sigmoid function

%and the loss function of $x$ is further calculated based on the reconstructed $x$.

After the loss functions of all the data in $n_{i,k}^{r}$ are calculated, the local loss function for iteration $k$ is calculated as
\begin{equation}
f(\omega_{i,k}^r)=\frac{1}{\left| n_{i,k}^r\right|}\sum_{x\in n_{i,k}^r} l\left(\omega_{i,k}^r;x\right),
\label{eq7}
\end{equation}
where $\left| n_{i,k}^r\right|$ is the number of data in $n_{i,k}^r$.

Then the regularized local loss function is calculated to reduce the deviation between the local model $\omega_{i,k}^r$ and global model $\omega^{r}$ to improve the algorithm convergence, i.e.,
\begin{equation}
g\left(\omega_{i,k}^r\right)=f\left(\omega_{i,k}^r\right)+\frac{\rho}{2}\left\|\omega^{r}-\omega_{i,k}^r\right\|^{2},
\label{eq8}
\end{equation}
where $\rho$ is the regularization parameter.

%A vehicle may not transmit in the previous round due to

Let $\nabla g(\omega_{i,k}^{r})$ be the gradient of $g\left(\omega_{i,k}^r\right)$, which is referred to as the local gradient. In the previous round, some vehicles may upload the updated local model unsuccessfully due to the delayed training time, and thus adversely affect the convergence of global model \cite{Chen2020}\cite{Xie2019}\cite{-S2021}. Here, these vehicles are called stragglers and the local gradient of a straggler in the previous round is referred to as the delayed local gradient. To solve this problem, the delayed local gradient will be aggregated into the local gradient of the current round $r$. Thus, the aggregated local gradient can be calculated as
\begin{equation}
\nabla \zeta_{i,k}^{r}=\nabla g(\omega_{i,k}^{r})+\beta \nabla g_{i}^{d},
\label{eq9}
\end{equation}
where $\beta$ is the decay coefficient and $\nabla g_{i}^{d}$ is the delayed local gradient. Note that $\nabla g_{i}^{d}=0$ if $V_{i}^{r}$ uploads successfully in the previous round.

Then the local model for the next iteration is updated as
\begin{equation}
\omega^{r}_{i,k+1}=\omega^{r}-\eta_{l}^{r}\nabla \zeta_{i,k}^{r},
\label{eq10}
\end{equation}where $\eta_{l}^{r}$ is the local learning rate in round $r$, which is calculated as
\begin{equation}
\eta_{l}^{r}=\eta_{l} \max \{1, \log (r)\},
\label{eq11}
\end{equation} where $\eta_{l}$ is the initial value of local learning rate.

Then iteration $k$ is finished and $V_{i}^{r}$ randomly samples some training data again to start the next iteration. When the number of iterations reaches the threshold $e$, $V_{i}^{r}$ completes the local training and upload the updated local model $\omega_{i}^{r}$ to the local RSU.

\subsubsection{Upload Model}
\
\newline
\indent
Each vehicle uploads its updated local model to the local RSU after it completes local training.

\subsubsection{Asynchronous aggregation}
\
\newline
\indent
If the local model of $V_{i}^{r}$, i.e., $\omega^{r}_{i}$, is the first model received by the local RSU, the upload is successful and the local RSU updates the global model. Otherwise, the local RSU drops $\omega^{r}_{i}$ and thus the upload is not successful.

When the upload is successful, the local RSU updates the global model $\omega^{r}$ by weighted averaging as follows:

%Considering the mobility of each VU and the transmission rate of V2R, we improve the weight of the global parameter aggregation. Then, we can rewrite the asynchronous aggregation method as follows:
\begin{algorithm}
	\caption{The Asynchronous Federated Learning Algorithm}
	\label{al1}
	Set global model $\omega^{r}$;\\
	\For{each round $r$ from $1$ to $R^{max}$}
  {
		\For{each vehicle $ V^{r}_{i} \in \mathbb{V}^{r}$ \textbf{in parallel}}
		{
			$T_{r,i}^{staying}=\left(L_{s}-P_{i}^{r}\right) / U_{i}^{r}$;\\
			\If{ $T_{r,i}^{staying}>T_{training}+T_{inference}$}
			{
			$V^{r}_i$ is selected to participate in asynchronous FL training;
			}
		}
		%\textbf{$C_{r}$}:the set of predicted popular contents in $ r$th round;\\
		%\textbf{$C_{i}$}:the set of predicted popular contents from VU $V_{r}^i$;\\
		\For{each selected vehicle $ V^{r}_{i}$}
		{
			$\omega^{r}_{i} \leftarrow \textbf{Vehicle Updates}(\omega^r,i)$;\\
			Upload the local model $\omega^{r}_{i}$ to the local RSU;\\
		}
		Receive the updated model $\omega^{r}_{i}$;\\
		Calculate the weight of the asynchronous aggregation $\chi_{i}$ based on Eq. \eqref{eq14};\\
		Update the global model based on Eq. \eqref{eq12};\\
	\Return $w^{r+1}$
	}
	\textbf{Vehicle Update}($w,i$):\\
	\textbf{Input:} $w^r$ \\
	Calculate the local learning rate $\eta_{l}^{r}$ based on Eq. \eqref{eq11};\\
	\For{each local epoch k from $1$ to $e$}
	{
		Randomly samples some data $n_{i,k}^r$ from the training set;\\
		\For{each data $x \in n_{i,k}^r$ }
		{
			Calculate the loss function of data $x$ based on Eq. \eqref{eq6};\\
		}
			Calculate the local loss function for interation $k$ based on Eq. \eqref{eq7};\\
			Calculate the regularized local loss function $g\left(\omega_{i,k}^r\right)$ based on Eq. \eqref{eq8};\\
			Aggregate local gradient $\nabla \zeta_{i,k}^{r}$ based on Eq. \eqref{eq9};\\
			Update the local model $\omega^{r}_{i,k}$ based on Eq. \eqref{eq10};\\
	}
	Set $\omega^{r}_{i}=\omega^{r}_{i,e}$;\\
	\Return$\omega^{r}_{i}$

\end{algorithm}

\begin{equation}
\omega^{r}=\omega^{r-1}+\frac{d_{i}^r}{d^r} \chi_{i} \omega^{r}_{i},
\label{eq12}
\end{equation}where $d_{i}^r$ is the size of local data in $V_i^r$, $d^r$ is the total local data size of the selected vehicles and $\chi_{i}$ is the weight of the asynchronous aggregation for $V_{i}^{r}$.
The weight of the asynchronous aggregation $\chi_{i}$ is calculated by considering the traversed distance of $V_{i}^{r}$ in the coverage area of the local RSU and the content transmission delay from local RSU to $V_{i}^{r}$ to improve the accuracy of the global model and reduce the content transmission delay. Specifically, if the traversed distance of $V_{i}^{r}$ is large, it may have long available time to participate in the training, thus its local model should occupy large weight for aggregation to improve the accuracy of the global model. In addition, the content transmission delay from local RSU to $V_{i}^{r}$ is important because $V_{i}^{r}$ would finally download the content from the local RSU when the content is either cached in the local or neighboring RSU. Thus, if the content transmission delay from local RSU to $V_{i}^{r}$ is small, its local model should also occupy large weight for aggregation to reduce the content transmission delay. The weight of the asynchronous aggregation $\chi_{i}$ is calculated as

\begin{equation}
\chi_{i}=\mu_{1} {(L_{s}-P_{i}^{r})}+\mu_{2} \frac{s}{R_{R, i}^{r}},
\label{eq13}
\end{equation}where $\mu_{1}$ and $\mu_{2}$ are coefficients of the position weight and transmission weight, respectively (i.e., $\mu_{1}+\mu_{2}=1$), $s$ is the size of each content. Thus, the content transmission delay from local RSU to $V_{i}^{r}$ is affected by the transmission rate between the local RSU and $V_{i}^{r}$, i.e., $R_{R, i}^{r}$. We can further calculate $\chi_{i}$ based on the normalized $L_{s}-P_{i}^{r}$ and $R_{R, i}^{r}$, i.e.,
\begin{equation}
\chi_{i}=\mu_{1} \frac{(L_{s}-P_{i}^{r})}{L_{s}}+\mu_{2} \frac{R_{R, i}^{r}}{\max _{k \in N^{r}}\left(R_{R, k}^{r}\right)}.
\label{eq14}
\end{equation}

Since the local RSU knows $dis(S,V_{i}^{r})$ and $P_{i}^{r}$ for each vehicle $i$ at the beginning of the asynchronous FL, the local RSU can calculate $R_{R, i}^{r}$ according to Eqs. \eqref{eq2} and \eqref{eq3}, and further calculate $\chi_{i}$ according to  Eq. \eqref{eq13}.

Up to now, the asynchronous FL in round $r$ is finished and the updated global model $\omega^{r}$ is obtained. The process of the asynchronous FL algorithm is shown in Algorithm \ref{al1} for ease of understanding, where $R^{max}$ is the maximum number of rounds, $e$ is the maximum number of local epochs. Then, the local RSU sends the obtained model to each vehicle to predict popular contents.

%After the training of the AE neural network, the hidden features of users and contents can be obtained which are used to predict its content popularity.

\subsection{Popular Content Prediction}

\begin{figure*}
\center
\includegraphics[scale=0.6]{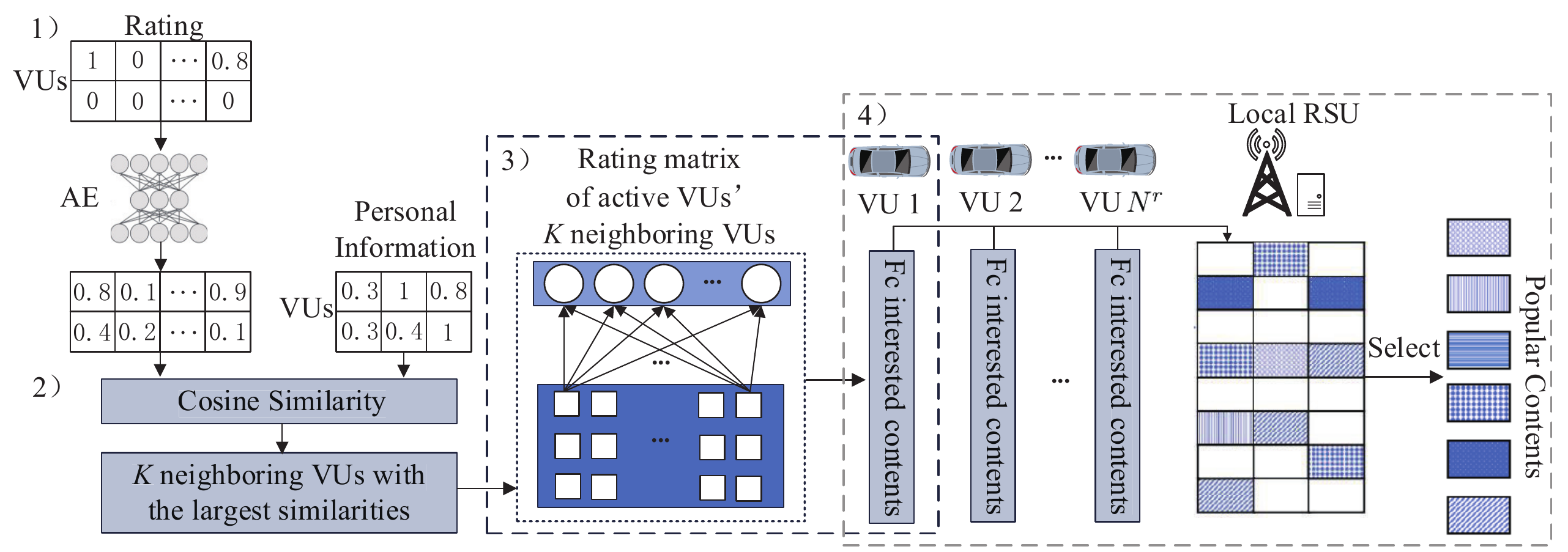}
\caption{Popular content prediction process}
\label{fig3}
\end{figure*}

In this subsection, we propose an algorithm to predict the popular contents. As shown in Fig. \ref{fig3}, the popular content prediction algorithm consists of the 4 steps as follows.

\subsubsection{Data Preprocessing}
\
\newline
\indent
The VU's rating for a content is $0$ when VU is uninterested in the content or has not requested a content. Thus, it is difficult to differentiate if a content is an interesting one for the VU when its rating is $0$. Marking all contents with rating $0$ as uninterested contents is a bias prediction. Therefore, we adopt the obtained model to reconstruct the rating for each content in the first step, which is described as follows.

Each vehicle abstracts a rating matrix from the data in the testing set, where the first dimension of the matrix is VUs' ID and the second dimension is VU's ratings for all contents. Denote the rating matrix of $V_{i}^r$ as $\boldsymbol{R}_{i}^r$. Then, the AE with the obtained model is adopted to reconstruct $\boldsymbol{R}_{i}^r$. The rating matrix $\boldsymbol{R}_{i}^r$ is used as the input data for the AE that outputs the reconstructed rating matrix $\hat{\boldsymbol{R}}_{i}^r$. Since $\hat{\boldsymbol{R}}_{i}^r$ is reconstructed based on the obtained model which reflects the hidden features of data, $\hat{\boldsymbol{R}}_{i}^r$ can be used to approximate the rating matrix $\boldsymbol{R}_{i}^r$.
%Then $R_{i,j}^r$ can be recovered from the latent code to generate the reconstructed rating matrix $\hat R_{i,j}^r$, $\hat R_{i,j}^r$ is a matrix with these unknown values.
% It is leveraged to extract the potential correlated features between the users and requested contents.
Then, similar to the rating matrix, each vehicle also abstracts a personal information matrix from the data of the testing set, where the first dimension of the matrix is VUs' ID and the second dimension is VU's personal information.

\subsubsection{Cosine Similarity}
\
\newline
\indent
$V_{i}^r$ counts the number of the nonzero ratings for each VU in $\boldsymbol{R}_{i}^r$ and marks the VUs with the $1$$/$$m$ largest numbers as active VUs. Then, each vehicle combines $\hat{\boldsymbol{R}}_{i}^r$ and the personal information matrix (denoted as $\boldsymbol{H}_{i}^r$) to calculate the similarity between each active VU and other VUs. The similarity between an active VU $a$ and $b$ is calculated according to cosine similarity \cite{yuet2018}
\begin{equation}
\begin{aligned}
\operatorname{sim}_{a,b}^{r,i}=\cos \left(\boldsymbol{H}_{i}^r(a,:), \boldsymbol{H}_{i}^r(b,:)\right)\\
=\frac{\boldsymbol{H}_{i}^r(a,:) \cdot \boldsymbol{H}_{i}^r(b,:)^T}{\left\|\boldsymbol{H}_{i}^r(a,:)\right\|_{2} \times\left\|\boldsymbol{H}_{i}^r(b,:)\right\|_{2}}
\label{eq15}
\end{aligned}
\end{equation}where $\boldsymbol{H}_{i}^r(a,:)$ and $\boldsymbol{H}_{i}^r(b,:)$ are the vectors corresponding to the active VU $a$ and $b$ in the combined matrixes, respectively, $\left\|\boldsymbol{H}_{i}^r(a,:)\right\|_{2}$ and $\left\|\boldsymbol{H}_{i}^r(b,:)\right\|_{2}$ are the 2-norm of $\boldsymbol{H}_{i}^r(a,:)$ and $\boldsymbol{H}_{i}^r(b,:)$, respectively. Then for each active VU $a$, $V_{i}^r$ selects the VUs with the $K$ largest similarities as the $K$ neighboring VUs of VU $a$. The ratings of the $K$ neighboring VUs also reflect the preferences of VU $a$ to a certain extent.

\subsubsection{Interested Contents}
\
\newline
\indent
After determining the neighboring VUs of active VUs, in $\boldsymbol{R}_{i}^r$, the vectors of neighboring VUs for each active VU are abstracted to construct a matrix $\boldsymbol{H}_K$, where the first dimension of $\boldsymbol{H}_K$ is the IDs of the neighboring VUs for active VUs, while the second dimension of $\boldsymbol{H}_K$ is the ratings of the contents from neighboring VUs. In $\boldsymbol{H}_K$, a content with a VU's nonzero rating is regarded as the VU's interested content. Then the number of interested contents is counted for each VU, where the counted number of a content is referred to as the content popularity of the content. $V_{i}^r$ selects the contents with the $F_c$ largest content popularity as the predicted interested contents.

\subsubsection{Popular Contents}
\
\newline
\indent
After vehicle in the local RSU uploads their predicted interested contents, the local RSU collects and compares the predicted interested contents uploaded from all vehicles to select the contents with the $F_{c}$ largest content popularity as the popular contents. The proposed popular content prediction algorithm is illustrated in Algorithm \ref{al2}, where $\mathbb{C}^{r}$ is the set of the popular contents and $\mathbb{C}_{i}^r$ is the set of interested contents of $V^{r}_i$.

\begin{algorithm}
	\caption{The Popular Content Prediction Algorithm}
	\label{al2}
		\textbf{Input: $\omega^{r}$}\\
		\For{each vehicle $ V^{r}_{i} \in \mathbb{V}^{r}$}
		{
			Construct the rating matrix $\boldsymbol{R}_{i}^r$ and personal information matrix;\\
			$\hat{\boldsymbol{ R}}_{i}^r \leftarrow AE(\omega^{r},\boldsymbol{R}_{i}^r)$;\\
			Combine $\hat{\boldsymbol{ R}}_{i}^r$ and information matrix as $\boldsymbol{H}_{i}^r$;\\
			$\mathbb{C}_{i}^r \leftarrow \textbf{Vehicle Predicts}(\boldsymbol{H}_{i}^r,i)$;\\
			Uploads $\mathbb{C}_{i}^r$ to the local RSU;\\
		}
		\textbf{Compare} received contents and select the $F_c$ most interested contents into $\mathbb{C}^{r}$.\\
	\Return $\mathbb{C}^{r}$\\
	\textbf{Vehicle Predicts}$(\boldsymbol{H}_{i}^r, i)$:\\
	\textbf{Input: $\boldsymbol{H}_{i}^r, i\in {1,2,...,N^r}$}\\
	Calculate the similarity between $V_{i}^r$ and other vehicles based on Eq. \eqref{eq15};\\
	Select the first $K$ vehicles with the largest similarity as neighboring vehicles of $V_{i}^r$;\\
	Construct reconstructed rating matrixes of $K$ neighboring vehicles as $\boldsymbol{H}_K$;\\
	Select the $F_c$ most interested contents as $\mathbb{C}_{i}^r$;\\
	\Return $\mathbb{C}_{i}^r$

\end{algorithm}

The cache capacity of the each RSU $c$, i.e., the largest number of contents that each RSU can accommodate, is usually smaller than $F_{c}$.
Next, we will propose a cooperative caching to determine where the predicted popular contents can be cached.

\subsection{Cooperative Caching Based on DRL}
%After deriving the $F_{c}$ popular contents, the next challenge is to determine which contents in $F_{c}$ popular contents should be proactively cached in the appropriate RSUs. To approach this target,

We consider the computation capability of each RSU is powerful and the cooperative caching can be determined within a short time. The main goal is to find an optimal cooperative caching based on DRL to minimize the content transmission delay. Next, we will formulate the DRL framework and then introduce the DRL algorithm.

%and DRL deciding content placement are two separate processes, i.e., RSU first saves information of requested contents and the predicted content popularity to its database, and when the data in the database reaches a certain amount, we perform DRL model updates. In order to ensure real-time VN scenario, the DRL model adopted by RSU is the previously trained model so that decisions can be made quickly. Next, we first formulate the DRL framework and then introduce the dueling deep Q-network (DQN) algorithm.
\subsubsection{DRL Framework}
\
\newline
\indent
The DRL framework includes state, action and reward. The training process is divided into slots. For the current slot $t$, the local RSU observes the current state $s(t)$ and decides the current action $a(t)$ based on $s(t)$ according to a policy $\pi$, which is used to generate the action based on the state at each slot. Then the local RSU can obtain the current reward $r(t)$ and observes the next state $s(t+1)$ that is transited from the current state $s(t)$. We will design $s(t)$, $a(t)$ and $r(t)$, respectively, for this DRL framework.

\paragraph{State}
\
\newline
\indent
We consider the contents cached by the local RSU as the current state $s(t)$. In order to focus on the contents with high popularity, the contents of the state space $s(t)$ are sorted in descending order based on the predicted content popularity of the $F_c$ popular contents, thus the current state can be expressed as $s(t)=\left(s_{1}, s_{2}, \ldots, s_{c}\right)$, where $s_{i}$ is the $i$th most popular content.

%We sort the content indexes of the state space $s(t)$ in descending order based on the $F_c$ predicted popular contents, which can reduce the frequency of low content popularity indexes appearing in $s(t)$.

%state $S=\left(s_{1}, s_{2}, \ldots, s_{c}\right)$, where $s(t)$ represents the specific cache contents of $S_i$ and $c$ is the cache capacity of the RSU. We sort the content indexes of the state space $s(t)$ in descending order based on the $F_c$ predicted popular contents, which can reduce the frequency of low content popularity indexes appearing in $s(t)$.
\paragraph{Action}
\
\newline
\indent
Action $a(t)$ represents whether the contents cached in the local RSU need to be relocated or not. In the $F_c$ predicted popular contents, the contents that are not cached in the local RSU form a set $\mathbb{N}$. If $a(t)=1$, the local RSU randomly selects $n(n<c)$ contents from $\mathbb{N}$ and exchanges them with the $n$ lowest popular contents cached in the local RSU, and then sorts the contents in a descending order based on their content popularity to get $s(t+1)$. Neighboring RSU also randomly samples $c$ contents from $F_c$ popular contents that do not belong to $s(t+1)$ as the cached contents of the neighboring RSU within the next slot $t+1$. We denote the contents cached by the neighboring RSU as $s_n(t+1)$.
If $a(t)=0$, the contents cached in the local RSU will not be relocated and the neighboring RSU also determines its cached contents, similar to the case when $a(t)=1$.

\paragraph{Reward}
\
\newline
\indent
The reward function $r(t)$ is designed to minimize the total content transmission delay to fetch the contents requested by vehicles. Note that the local RSU has recorded all the contents requested by the vehicles. The content transmission delays to fetch a requested content $f$ are different when the content is cached in different places.

If content $f$ is cached in the local RSU, i.e., $f\in s(t)$,  the local RSU transmits content $f$ to $V_{i}^{r}$, thus the content transmission delay is calculated as

\begin{equation}
d_{R, i, f}^{r}=\frac{s}{R_{R, i}^{r}},
\label{eq16}
\end{equation}where $R_{R, i}^{r}$ is the transmission rate between the local RSU and $V_{i}^{r}$, which has been calculated by Eq. \eqref{eq3}.

If content $f$ is cached in the neighboring RSU, i.e., $f\in s_n(t)$, the neighboring RSU sends the content to the local RSU that forwards the content to $V_{i}^{r}$, thus the transmission delay is calculated as

\begin{equation}
\bar{d}_{R, i, f}^{r}=\frac{s}{R_{R, i}^{r}}+\frac{s}{R_{R-R}},
\label{eq17}
\end{equation}where $R_{R-R}$ is the transmission rate between the local RSU and neighboring RSU, which is a constant transmission rate in the wired link.

If content $f$ is neither cached in the local RSU nor in the neighboring RSU, i.e., $f \notin s(t) \text{ and } f \notin s_n(t)$, the MBS transmits content $f$ to $V_{i}^{r}$, thus the content transmission delay is expressed as

\begin{equation}
d_{B, i,f}^{r}=\frac{s}{R_{B, i}^{r}},
\label{eq18}
\end{equation}where $R_{B, i}^{r}$ is the transmission rate between the MBS and $V_{i}^{r}$, which is calculated according to Eq. \eqref{eq4}.

In order to clearly distinguish the content transmission delays under different conditions, we set the reward that $V_{i}^r$ fetches content $f$ at slot $t$  as

\begin{equation}
r_{i,f}^r(t)=\begin{cases}
 e^{-\lambda_{1} d_{R,i,f}^{r}}& f\in s(t)\\
 e^{-\left(\lambda_{1} d_{R, i, f}^{r}+\lambda_{2} \bar d_{R, i, f}^{r}\right)}&f \in s_n(t) \\
 e^{-\lambda_{3} d_{M, i, f}^{r}}&f \notin s(t) \text{ and } f \notin s_n(t)
 \end{cases},
\label{eq19}
\end{equation}
where $\lambda_{1}+\lambda_{2}+\lambda_{3}=1$ and $\lambda_{1}<\lambda_{2}\ll \lambda_{3}$.

Thus the reward function $r(t)$ is calculated as
\begin{equation}
r(t)=\sum_{i=1}^{N^r}\sum_{f=1}^{F_{i}^r} r_{i,f}^r(t),
\label{eq20}
\end{equation}where $F_{i}^r$ is the number of requested contents from $V_{i}^r$.

%Specifically, if the requested content $f$ from $V_{i}^r$ cached in $s(t)$, it means that $V_{i}^r$ can fetch $f$ from the local RSU and in this case the value of $r_{i,f}^r(t)$ is $e^{-\lambda_{1} d_{R,i,f}^{r}}$. If the requested content $f$ from $V_{i}^r$ cached in the neighbouring RSU $s_n(t)$, it means that $V_{i}^r$ can fetch content $f$ from the neighbouring RSU and in this case the value of $r_{i,f}^r(t)$ is $e^{-\left(\lambda_{1} d_{R, i, f}^{r}+\lambda_{2} \bar d_{R, i, f}^{r}\right)}$. If neither of the above conditions is met for slot $t$, it means that $V_{i}^r$ can fetch content $f$ from MBS and in this case the value of $r_{i,f}^r(t)$ is $e^{-\lambda_{3} d_{M, i, f}^{r}}$,
%where $\lambda_{1}+\lambda_{2}+\lambda_{3}=1$, $\lambda_{1}<\lambda_{2}\ll \lambda_{3}$.

\subsubsection{DRL Algorithm}
\
\newline
\indent
As mentioned above, the next state will change when the action is $1$. The dueling DQN algorithm is a particular algorithm which works for the cases where the partial actions have no relevant effects on subsequent states \cite{Wangarxiv2016}. Specifically, the dueling DQN decomposes the Q-value into two functions $V$ and $A$. Function $V$ is the state value function that is unrelated to the action, while $A$ is the action advantage function that is related to the action. Therefore, we adopt the dueling DQN algorithm to solve this problem.

\begin{algorithm}
	\caption{Cooperative Caching Based on Dueling DQN Algorithm}
	\label{al3}
	Initialize replay buffer $\mathcal{D}$, the parameters of the prediction network $\theta$, the parameters of the target network $\theta'$;\\
	\textbf{Input:} requested contents from all vehicles in the local RSU for round $r$\\
	\For{episode from $1$ to $T_s$}
	{
			Local RSU randomly caches $c$ contents from $F_c$ popular contents;\\
			Neighboring RSU randomly caches $c$ contents from $F_c$ popular contents that are not cached in the local RSU;\\
		\For{slot from $1$ to $N_s$}
{
			Observe the state $s(t);$\\
			Calculate the Q-value of prediction network $Q(s(t), a; \theta)$ based on Eq. \eqref{eq21};\\
			Calculate the action $a(t)$ based on Eq. \eqref{eq22};\\
			Obtain state $s(t+1)$ after executing action $a(t)$;\\
			Obtain reward $r(t)$ based on Eqs. \eqref{eq16} - \eqref{eq20};\\
			Store tuple $(s(t),a(t),r(t),s(t+1))$ in $\mathcal{D}$;\\
			\If{number of tuples in $\mathcal{D}$ is larger than $I$}
{
			Randomly sample a minibatch of $I$ tuples from $\mathcal{D}$;\\
			\For{tuple $i$ from $1$ to $I$}
{
			Calculate the Q-value function of target network $Q'(s^i, a; \theta')$ based on Eq. \eqref{eq23};\\
			Calculate the target Q-value of the target network $y^i$ based on Eq. \eqref{eq24};\\
			Calculate the loss function $L(\theta)$ based on Eq. \eqref{eq25};\\
}
			Calculate the gradient of loss function $\nabla_{\theta} L(\theta)$ based on Eq. \eqref{eq26};\\
			Update parameters of the prediction network $\theta$ based on Eq. \eqref{eq27};\\
}
\If{number of slots is $M$}
{$\theta'=\theta$.\\}
}
	}
\end{algorithm}
%, which is the average reward of executing action $a(t)$ to solve the problem of reward bias under state $s(t)$

The dueling DQN includes a prediction network, a target network and a replay buffer. The prediction network evaluates the current state-action value (Q-value) function, while the target network generates the optimal Q-value function. Each of them consists of three layers, i.e., the feature layer, the state-value layer, and the advantage layer. The replay buffer $\mathcal{D}$ is adopted to cache the transitions for each slot. The dueling DQN algorithm is illustrated in Algorithm \ref{al3} and is described in detail as follow.

\begin{figure*}
\center
\includegraphics[scale=0.27]{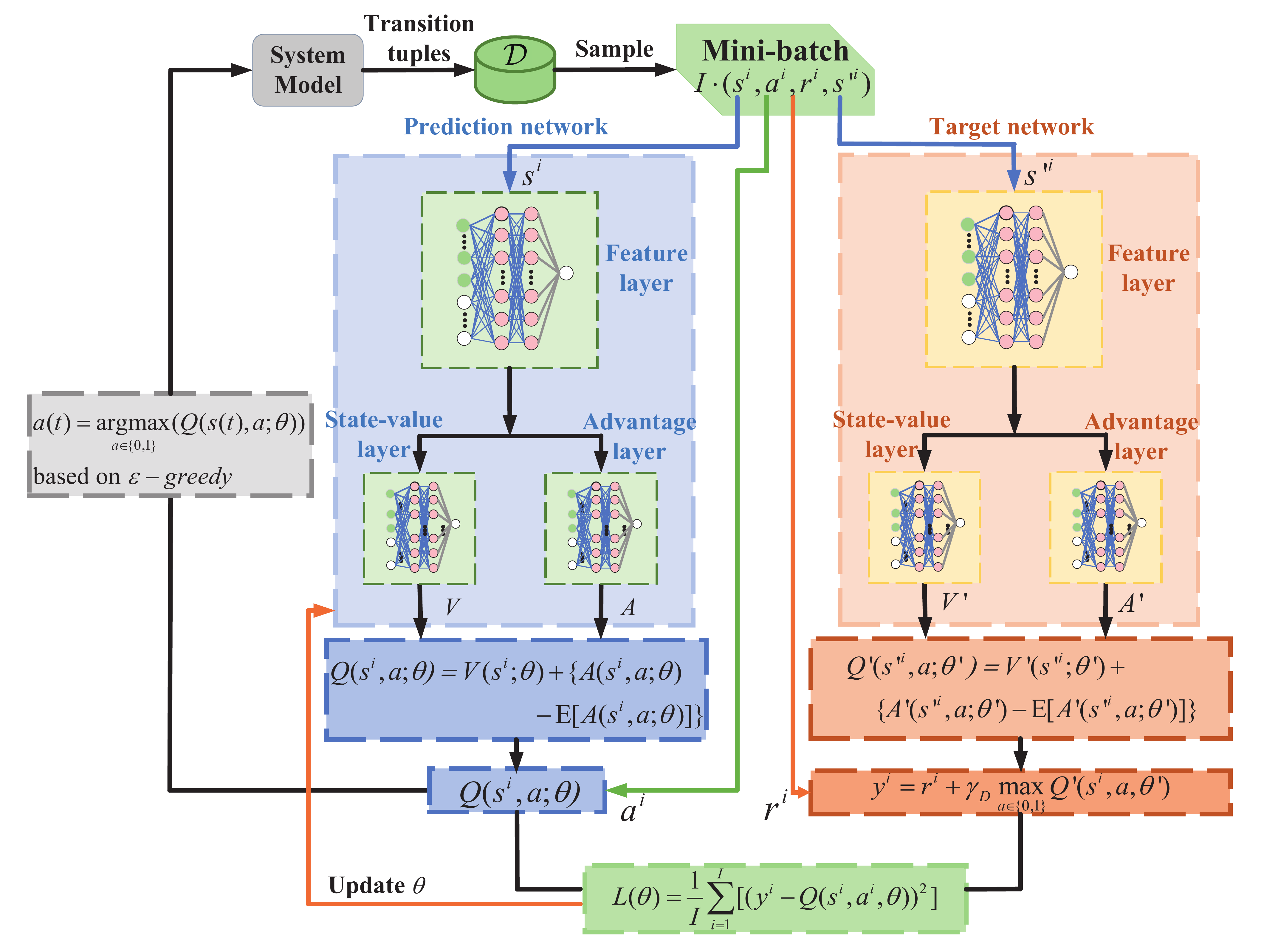}
\caption{The flow diagram of the dueling DQN}
\label{fig4}
\end{figure*}

Firstly, the parameters of the prediction network $\theta$ and the parameters of the target network $\theta'$ are initialized randomly. The requested contents from all vehicles in the local RSU for round $r$ as input (lines 1-2).

Then the algorithm is executed for $T_s$ episodes. At the beginning of each episode, the local RSU randomly selects $c$ contents from $F_c$ popular contents, and the neighboring RSU randomly selects $c$ contents from $F_c$ popular contents that are not cached in the local RSU. Then the algorithm is executed iteratively from slots $1$ to $N_s$. In each slot $t$, the local RSU first observes state $s(t)$ and then input $s(t)$ to the prediction network, in which it goes through the feature layer, state-value layer and advantage layer, respectively. In the end, the prediction network outputs the state value function $V(s(t) ; \theta)$ and the action advantage function under each action $a$, i.e., $A(s(t), a ; \theta)$, respectively, where ${a \in\{0,1\}}$. Furthermore, the Q-value function of prediction network under each action $a$ is calculated as
\begin{equation}
\begin{aligned}
Q(s(t), a; \theta)=V(s(t) ; \theta)+\{ A(s(t), a ; \theta) \\
-\mathbb{E}[A(s(t), a ; \theta)] \} \\
\end{aligned}.
\label{eq21}
\end{equation}

In Eq. \eqref{eq21}, the range of Q-values can be narrowed to remove redundant degrees of freedom by calculating the difference between the action advantage function $A(s(t), a ; \theta)$ and the average value of the action advantage functions under all actions, i.e., $\mathbb{E}[A(s(t), a ; \theta)]$. Thus, the stability of the algorithm can be improved.

%In practice, the action advantage function is generally calculated by the action advantage function under action $a$, i.e., $A(s(t), a ; \theta)$, minuses the average value of the action advantage functions under all actions, i.e., $\mathbb{E}[A(s(t), a ; \theta)]$. This ensures the range of Q-values can be narrowed to remove redundant degrees of freedom and improve the stability of the algorithm.

Then action $a(t)$ is chosen by the $\varepsilon \text {-greedy}$ method, which is calculated as
\begin{equation}
a(t)=\underset{a \in\{0,1\}}{\operatorname{argmax}}(Q(s(t), a;\theta))
\label{eq22}.
\end{equation}
Particularly, action $a(1)$ is initialized as $1$ at slot $1$.

The local RSU calculates the reward $r(t)$ according to Eqs. \eqref{eq16} - \eqref{eq20} and state $s(t)$ transits to the next state $s(t+1)$, then the local RSU observes $s(t+1)$. Next, the neighboring RSU randomly samples $c$ popular contents that are not cached in $s(t+1)$ as its cached contents, which is denoted as $s_n(t+1)$. The transition from $s(t)$ to $s(t+1)$ is denoted as tuple $(s(t),a(t),r(t),s(t+1))$, which is then stored in the replay buffer $\mathcal{D}$. When the number of the stored tuples in the replay buffer $\mathcal{D}$ is larger than $I$, the local RSU randomly samples $I$ tuples from $\mathcal{D}$ to form a minibatch. Let $(s^i,a^i,r^i,s'^i), (i=1,2,3,...,I)$ be the $i$-th tuple in the mini-batch. Then $S_i$ input each tuple into the prediction network and the target network (lines 3-12).

Next, we will introduce how parameters of prediction network $\theta$ are updated. For tuple $i$, the local RSU inputs $s^i$ into the target network, where it goes through the feature layer and outputs its feature. Then the feature is input to the state-value layer and the advantage layer, respectively, which output state value function $V'(s^i ; \theta')$ and action advantage function $A'(s^i, a; \theta')$ under each action $a \in \{0,1\}$, respectively. Thus, the Q-value function of target network of tuple $i$ under each action $a$ is calculated as
\begin{equation}
\begin{aligned}
&Q'(s^i, a; \theta')=V'(s^i ; \theta')\\
&+\{ A'(s^i, a ; \theta') -\left.\mathbb{E}\left[A'\left(s^i, a ; \theta'\right)\right]\right\}\\
\end{aligned}.
\label{eq23}
\end{equation}

Then the target Q-value of the target network of tuple $i$ is calculated as
\begin{equation}
y^i=r^i+\gamma_{D} \max _{a\in\{0,1\} } Q'(s^i, a; \theta'),
\label{eq24}
\end{equation}where $\gamma_{D}$ is the discount factor. The loss function is calculated as follows
\begin{equation}
L(\theta)=\frac{1}{I} \sum_{i=1}^{I}\left[(y^i-Q(s^i, a^i, \theta))^{2}\right].
\label{eq25}
\end{equation}

The gradient of loss function $\nabla_{\theta} L(\theta)$ for all sampled tuples is calculated as
\begin{equation}
\nabla_{\theta} L(\theta)=\frac{1}{I} \sum_{i=1}^{I} [\left(y^i-Q(s^i, a^i, \theta)\right) \nabla_{\theta^i} Q(s^i, a^i, \theta)].
\label{eq26}
\end{equation}

At the end of slot $t$, the parameters of the prediction network $\theta$ are updated as
\begin{equation}
\theta \leftarrow \theta-\eta_{\theta} \nabla_{\theta} L(\theta),
\label{eq27}
\end{equation}where $\eta_{\theta}$ is the learning rate of prediction network.

Up to now, the iteration in slot $t$ is completed, which will be repeated. During the iterations, the parameters of target network $\theta' $ are updated after a certain number of slots ($M$), as the parameters of prediction network $\theta$. When the number of slots reaches $N_s$, this episode is finished and then the local RSU randomly caches $c$ contents from $F_c$ popular contents to start the next episode. When the number of episodes reaches $T_s$, the algorithm will be terminated (lines 13-22). The flow diagram of the dueling DQN algorithm is shown in Fig. \ref{fig4}.

Finally, the local RSU and neighboring RSU cache popular contents according to the optimal cooperative caching, and then each vehicle fetches contents from the VEC. This round is finished after each vehicle has fetched contents and then the next round is started.

\section{Simulation and Analytical Results}
\label{sec6}

\begin{table}
\caption{Values of the parameters in the experiments.}
\label{tab2}
\footnotesize
\centering
\begin{tabular}{|c|c|c|c|}
  \hline
  \multicolumn{4}{|c|}{Parameters of System Model}\\
  \hline
  \textbf{Parameter} &\textbf{Value} &\textbf{Parameter} &\textbf{Value}\\
  \hline
	$B$ & $540$ kHz & $K$ &$10$\\
	\hline
  $m$ &$3$ & $p_B$  & $30$ dBm\\
	\hline
	$p_M$ & $43$ dBm & $R_{R,R}$ & $15$ Mbps \\
  \hline
  $s$ &$100$ bytes & $T_{training}$ & $2$s\\
  \hline
	$T_{inference}$ & $0.5$s & $U_{\max}$ &$60$ km/h\\
  \hline
  $U_{\min }$ &$50$ km/h & $\mu$ &$55$ km/h\\
  \hline
  $\sigma$ &$2.5$km/h & $\sigma_{c}^{2}$ & $-114$ dBm\\
  \hline
	\multicolumn{4}{|c|}{Parameters of Asynchronous FL}\\
  \hline
  \textbf{Parameter} &\textbf{Value} &\textbf{Parameter} &\textbf{Value}\\
  \hline
  $L_s$ &$1000$m & $\beta$ & $0.001$\\
  \hline
	$\eta_{l}$ &$0.01$ & $\mu_{1}$ &$0.5$ \\
	\hline
  $\mu_{2}$ &$0.5$ & $\rho$ &$0.0001$\\
  \hline
  \multicolumn{4}{|c|}{Parameters of DRL}\\
  \hline
  \textbf{Parameter} &\textbf{Value} &\textbf{Parameter} &\textbf{Value}\\
  \hline
  $I$ &$32$ & $\gamma_{D}$ & $0.99$\\
  \hline
	$\eta_{\theta}$ &$0.01$ & $\lambda_{1}$ & $0.0001$\\
  \hline
  $\lambda_{2}$ & $0.4$ & $\lambda_{3}$ & $0.5999$\\
  \hline
\end{tabular}
\end{table}

We have evaluated the performance of the proposed CAFR scheme in this section.
\subsection{Settings and Dataset}
We simulate a VEC environment on the urban road as shown in Fig. \ref{fig1} and the simulation tool is Python $3.8$. The communications between vehicle and RSU/MBS employ the 3rd Generation Partnership Project (3GPP) cellular V2X (C-V2X) architecture, where the parameters are set according to the 3GPP standard \cite{3gpp}. The simulation parameters are listed in Table \ref{tab2}. A real-world dataset from the MovieLens website, i.e., MovieLens 1M, is used in the experiments. MovieLens 1M contains $1,000,209$ rating values for $3,883$ movies from $6,040$ anonymous VUs with movie ratings ranging from $0$ to $1$, where each VU rates at least $20$ movies \cite{Harper2016}. MovieLens lM also provides personal information about VUs including ID number, gender, age and postcode. We randomly divide MovieLens lM data set to each vehicle as its local data. Each vehicle randomly chooses $99.8\%$ data from its local data as its training set and $0.2\%$ data as its testing set. For each round, each vehicle randomly samples a part of the movies from testing set as its requested contents.

%We divide MovieLens lM data set into a training set and a testing set, where 99.8\% is the training set and 0.2\% is the testing set, and then, for each round, the data are assigned randomly from the training set of each vehicle to form its local data and a part of the movies that have been rated by VUs are sampled randomly from the testing set of each vehicle as the vehicle's requested contents.

%the data from the training set randomly assigned to each vehicle as its local data and the data from the the testing set randomly sampled a part of the movies have been rated by users as the requested contents.

\subsection{Performance Evaluation}
We use cache hit ratio and the content transmission delay as performance metrics to evaluate the CAFR scheme. The cache hit rate is defined as the probability of fetching requested contents from the local RSU \cite{Muller2017}. If a requested content is cached in the local RSU, it can be fetched directly from the local RSU, which is referred to as a cache hit, otherwise, it is referred to as a cache miss. Thus, the cache hit rate is calculated as
\begin{equation}
\text { cache hit radio }=\frac{\text {cache hits }}{\text {cache hits }+\text {cache misses }}\times 100\%.
\label{eq28}
\end{equation}

The content transmission delay indicates the average delay for all vehicles to fetch contents, which is calculated as

\begin{equation}
\text {content transmission delay}=\frac{D^{\text {total}}}{\text {the number of vehicles }},
\label{eq29}
\end{equation}
where $D^{\text {total}}$ is the delay for all vehicles to fetch contents, and it is calculated by aggregating the content transmission delay for every vehicle to fetch contents.

We compare the CAFR scheme with other baseline schemes such as:
\begin{itemize}
\item Random: Randomly selecting $c$ contents from the all contents to cache in the local and neighboring RSU.
\item c-$\epsilon$-greedy: Selecting the contents with $c$ largest numbers of requests based on probability $1-\epsilon$ and selecting $c$ contents randomly based on probability $\epsilon$ to cache in the local RSU. In our simulation, $\epsilon= 0.1$.
\item Thompson sampling: For each round, the contents cached in the local RSU is updated based on the number of cache hits and cache misses in the previous round \cite{Cui2020}, and $c$ contents with the highest value are selected to cache in the local RSU.
\item FedAVG: Federated averaging (FedAVG) is a typical synchronous FL scheme where the local RSU needs to wait for the local model updates to update its global model according to weighted average method:
\begin{equation}
\omega^{r}=\sum_{i=1}^{N^r} \frac {d^r_i}{d^r} \omega^{r}_{i}.
\label{eq30}
\end{equation}

\item CAFR without DRL: Compared with the CAFR scheme, this scheme does not adopt the DRL algorithm to optimize caching scheme. Specifically, after predicting the popular contents, $c$ contents are randomly selected from the predicted popular contents to cache in the local RSU and neighboring RSU, respectively.
\end{itemize}

\begin{figure}
\center
\includegraphics[scale=0.5]{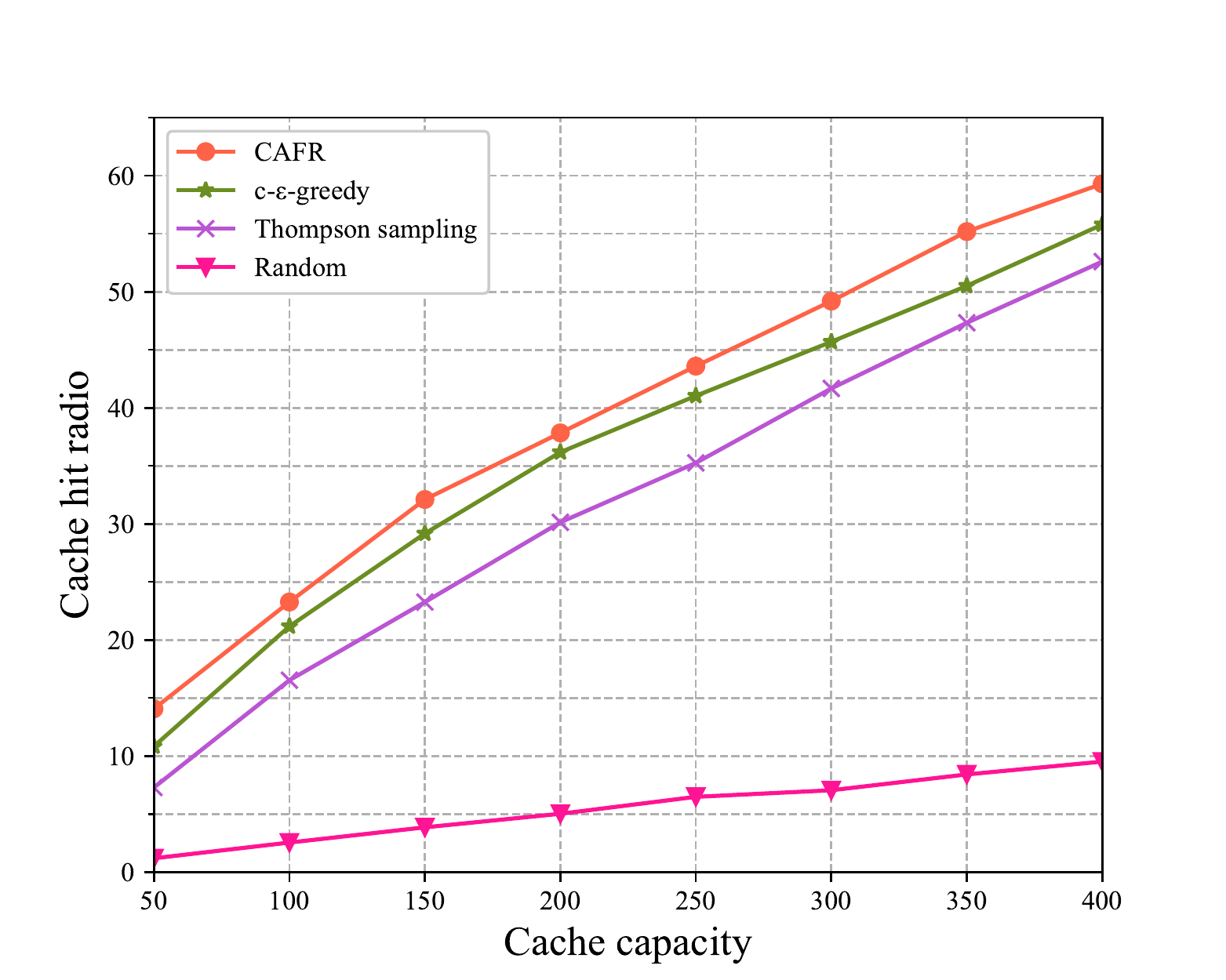}
\caption{Cache hit radio under different cache capacities}
\label{fig5}
\end{figure}

Now, we will evaluate the performance of the CAFR scheme through simulation experiments. In the following performance evaluation, each result is the average value of five experiments.

Fig. \ref{fig5} shows the cache hit ratio of different schemes under different cache capacities of each RSU, where the result of CAFR is obtained when the vehicle density is $15$ vehicles/km (i.e., the number of vehicles is 15 per kilometer), and the results of other schemes are independent with the vehicle density. It can be seen that the cache hit ratio of all schemes increases with a larger capacity. This is because that the local RSU caches more contents with a larger capacity, thus the requested contents of vehicles are more likely to be fetched from the local RSU. Moreover, it is seen that the random scheme provides the worst cache hit ratio, because the scheme just selects contents randomly without considering the content popularity. In addition, CAFR and c-$\epsilon$-greedy outperform the random scheme and the thompson sampling. This is because that random and thompson sampling schemes do not predict the caching contents through learning, whereas CAFR and c-$\epsilon$-greedy decide the caching contents by observing the historical requested contents. Furthermore, CAFR outperforms c-$\epsilon$-greedy. This is because that CAFR captures useful hidden features from the data to predict the accurate popular contents.

\begin{figure}
\center
\includegraphics[scale=0.5]{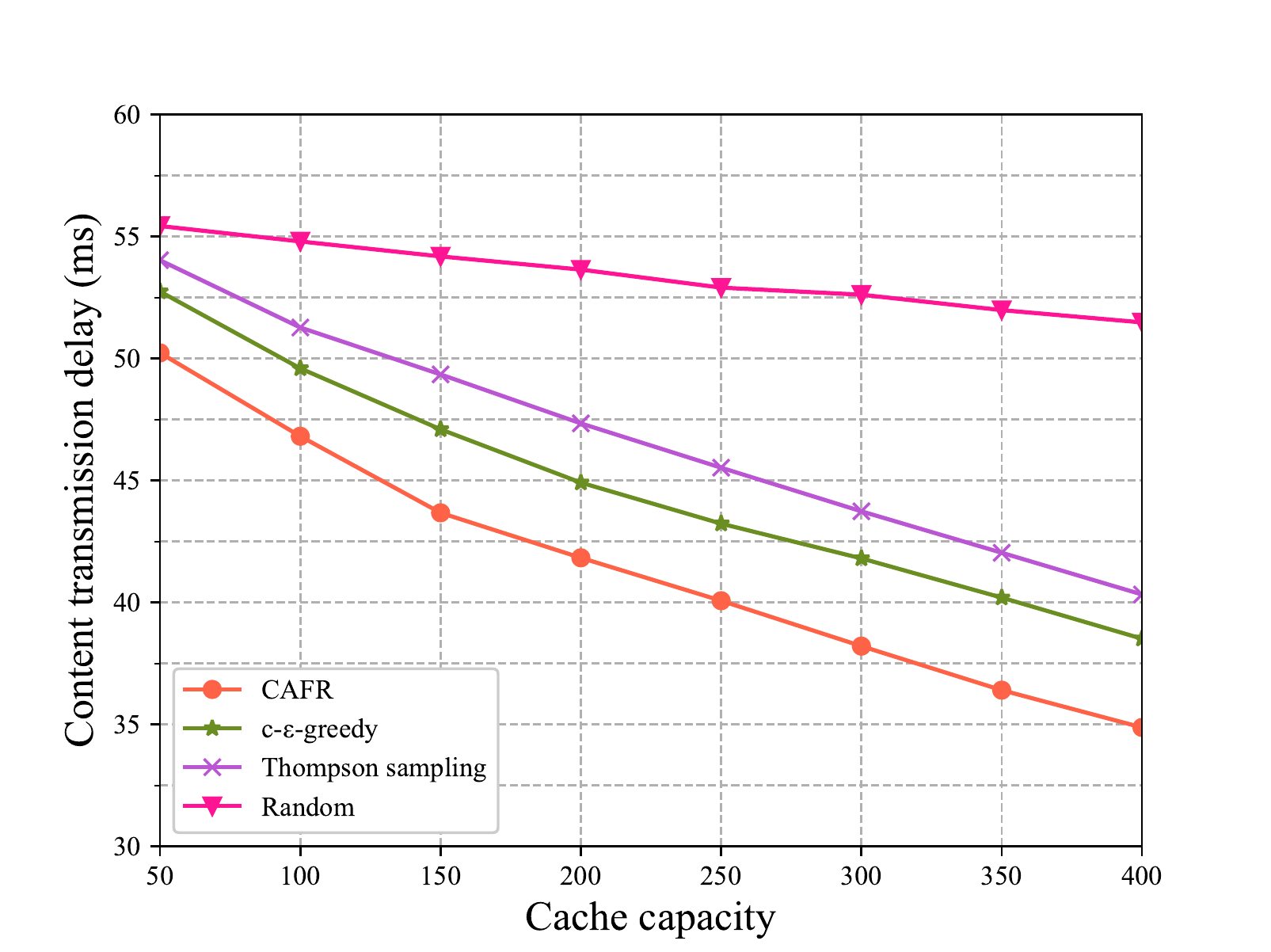}
\caption{Content transmission delay under different cache capacities}
\label{fig6}
\end{figure}

Fig. \ref{fig6} shows the content transmission delay of different schemes under different cache capacities of each RSU, where the vehicle density is $15$ vehicles/km. It is seen that the content transmission delays of all schemes decrease as the cache capacity increases. This is because that each RSU caches more contents as the cache capacity increases, and each vehicle fetches contents from local RSU and neighboring RSU with a higher possibility, thus reducing the content transmission delay. Moreover, the content transmission delay of CAFR is smaller than other schemes. This is because that the cache hit rate of CAFR is better than those of schemes, and more vehicles can fetch contents from local RSU directly, thus reducing the content transmission delay.

\begin{figure}
\center
\includegraphics[scale=0.5]{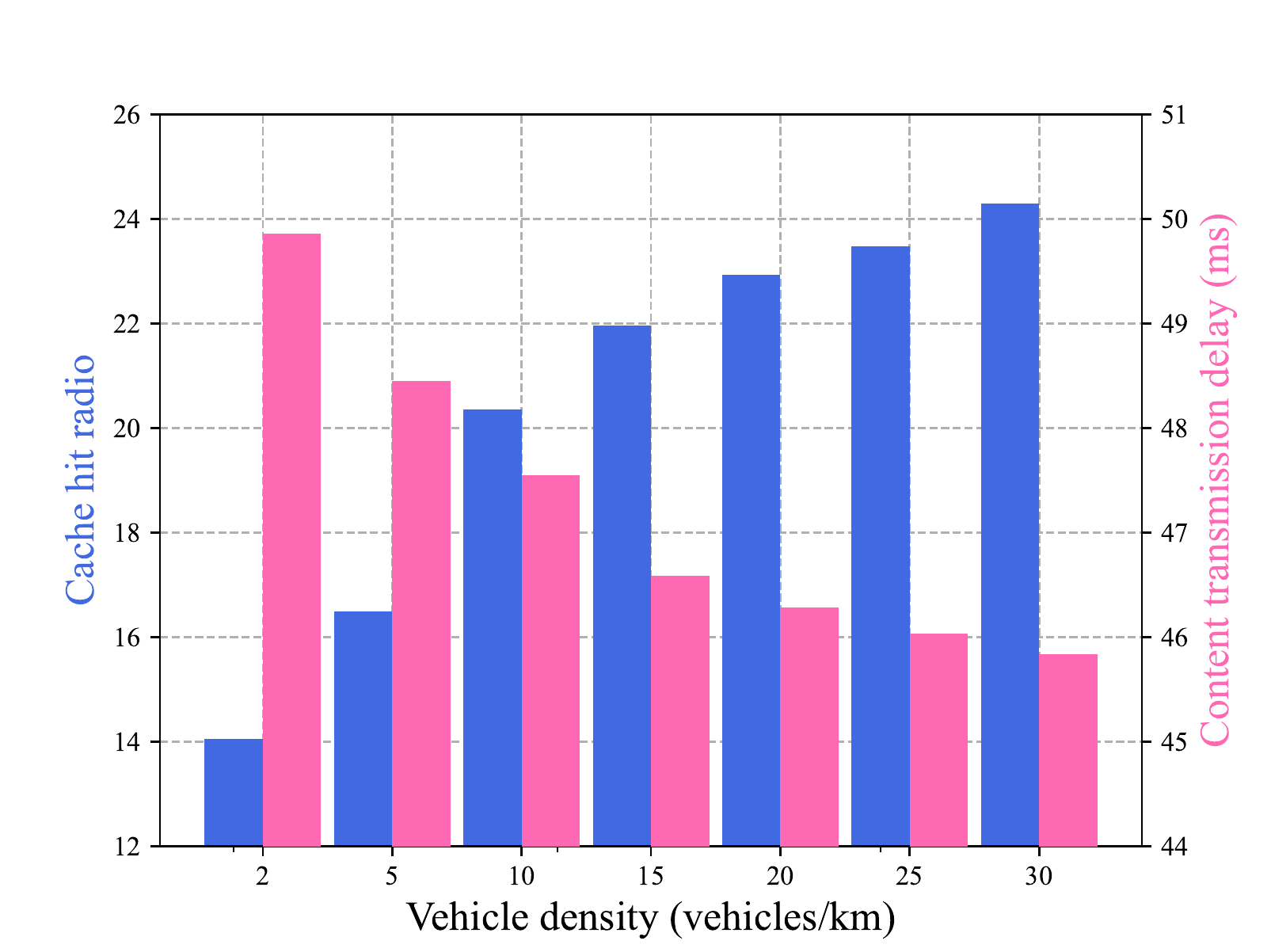}
\caption{Cache hit radio and content transmission delay under different vehicle densities}
\label{fig7}
\end{figure}

%Combined with Fig. \ref{fig6}, the higher the cache hit rate, the more chance the vehicle has to fetch contents from the local RSU, thus reducing content transmission delay. Since the cache hit rate of CAFR is better than other schemes, the content transmission delay of CAFR is also smaller than other schemes.

Fig. \ref{fig7} shows the cache hit ratio and the content transmission delay of the CAFR scheme under different vehicle densities when the cache capacity of each RSU is $100$. As shown in this figure, the cache hit rate increases as the vehicle density increases. This is because when more vehicles enter the coverage area of the RSU, the global model of the local RSU is trained based on more data, and thus can predict accurately. In addition, the content transmission delay decreases as the vehicle density increases. This is because the cache hit rate increases when the vehicle density increases, which enables more vehicles to fetch contents directly from local RSU.

%Combined with Fig. \ref{fig8}, as the vehicle density within the coverage area of RSU becomes larger, the cache hit rate is higher and the average content transmission delay is lower. From Fig. \ref{fig9}, it can be seen that the average content transmission delay from each vehicle in the coverage area of RSU is $308.15$ ms when the vehicle density is $2$ vehicles/km, however, the average content transmission delay from each vehicle within the coverage area of RSU is $145.58$ ms when the vehicle density is $5$ vehicles/km. With the increase of vehicle density, the average content transmission delay from each vehicle is decrease until $20.76$ ms when the vehicle density is $30$ vehicles/km.

%\begin{figure}
%\center
%\includegraphics[scale=0.5]{figure/ce_vs_round.eps}
%\caption{Cache hit radio and training time for communication rounds}
%\label{fig10}
%\end{figure}

\begin{figure}
\center
\includegraphics[scale=0.5]{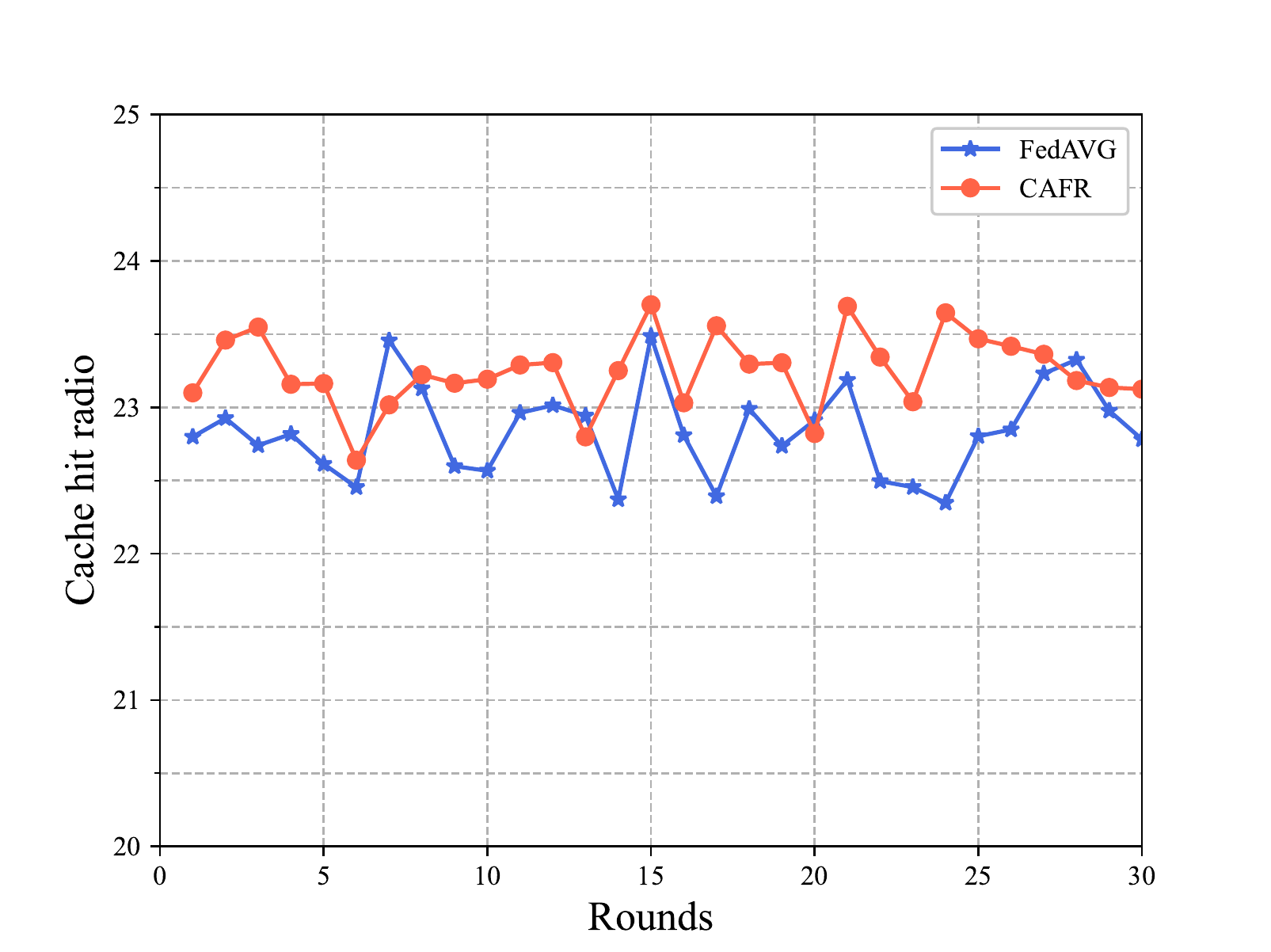}
\caption{Cache hit radio of CAFR and FedAVG}
\label{fig8}
\end{figure}

Fig. \ref{fig8} compares the cache hit rate of the CAFR scheme and the FedAVG scheme under different rounds when the vehicle density is $15$ vehicles/km and the cache capacity of each RSU is $100$ contents. It can be seen that the cache hit radio of CAFR fluctuates between $22.5\%$ and $24\%$ within $30$ rounds, while the cache hit rate of FedAVG scheme fluctuates between $22\%$ and $23.5\%$ within $30$ rounds. This indicates that the CAFR scheme is slightly better than the FedAVG scheme. This is because the CAFR scheme has considered the vehicles' mobility characteristics including the positions and velocities to select vehicles and aggregate the local model, thus improving the accuracy of the global model.

\begin{figure}
\center
\includegraphics[scale=0.5]{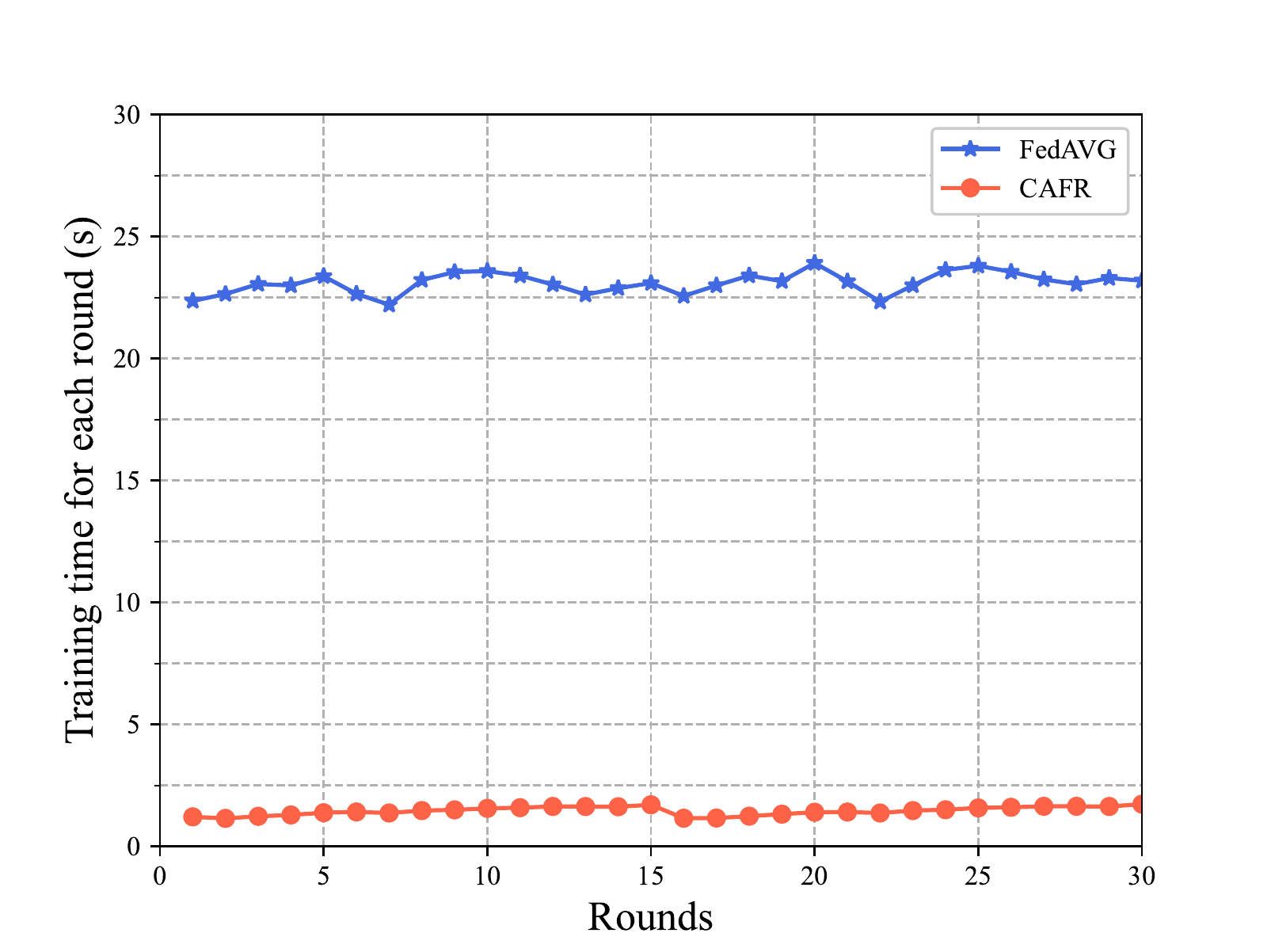}
\caption{Training time of CAFR and FedAVG}
\label{fig9}
\end{figure}

Fig. \ref{fig9} shows the training time of CAFR and FedAVG schemes for each round when the vehicle density is $15$ vehicles/km and the cache capacity of each RSU is $100$ contents. It can be seen that the training time of CAFR scheme for each round is within $1$s and $2$s, while the training time of FedAVG scheme for each round is within $22$s and $24$s. This indicates that CAFR scheme has a much smaller training time than the FedAVG scheme. This is because the FedAVG scheme needs to aggregate all vehicles' local models for the global model updating in each round, while the CAFR scheme aggregates as soon as a vehicle's local model is received for each round.

%Combined with Figs.\ref{fig11}, \ref{fig12}, we can conclude that compared to synchronous FL algorithm, our proposed MCFR algorithm has less training time per communication round and higher cache performance.
\begin{figure}
\center
\includegraphics[scale=0.5]{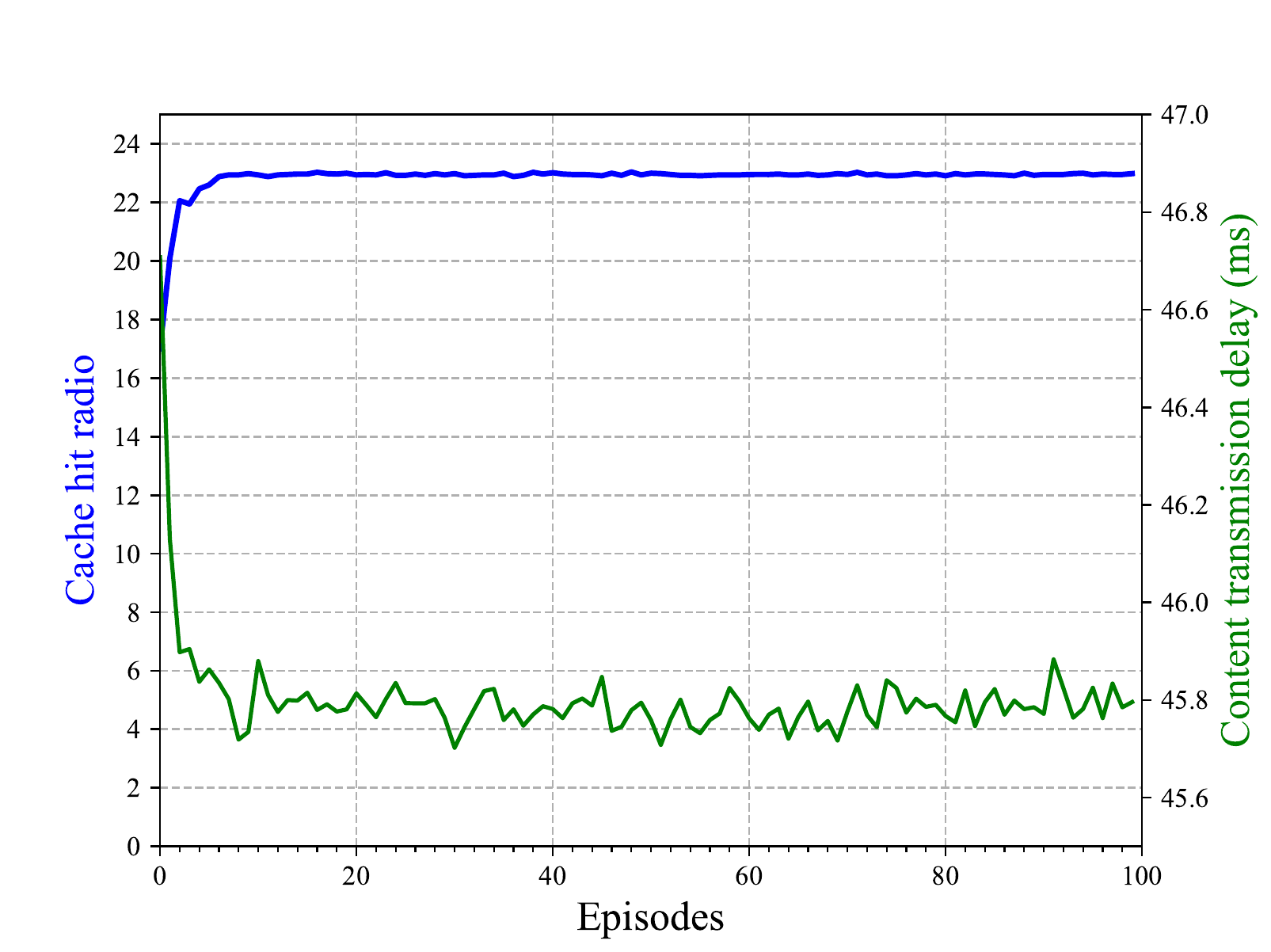}
\caption{Cache hit radio and content transmission delay of each episode in the DRL}
\label{fig10}
\end{figure}
Fig. \ref{fig10} shows the cache hit rate and content transmission delay of each episode in the DRL of the CAFR scheme when the vehicle density is $15$ vehicles/km and the cache capacity of RSU is $100$. As the episode increases, the cache hit rate gradually increases and the content transmission delay decreases gradually in the first ten episodes. This is because the local RSU and neighboring RSU gradually cache appropriate popular contents in the first ten episodes. In addition, it is seen that the cache hit rate and content transmission delay converge at around episode $10$. This is because the local RSU is able to learn the policy to perform optimal cooperative caching at around $10$ episodes.

\begin{figure}
\center
\includegraphics[scale=0.5]{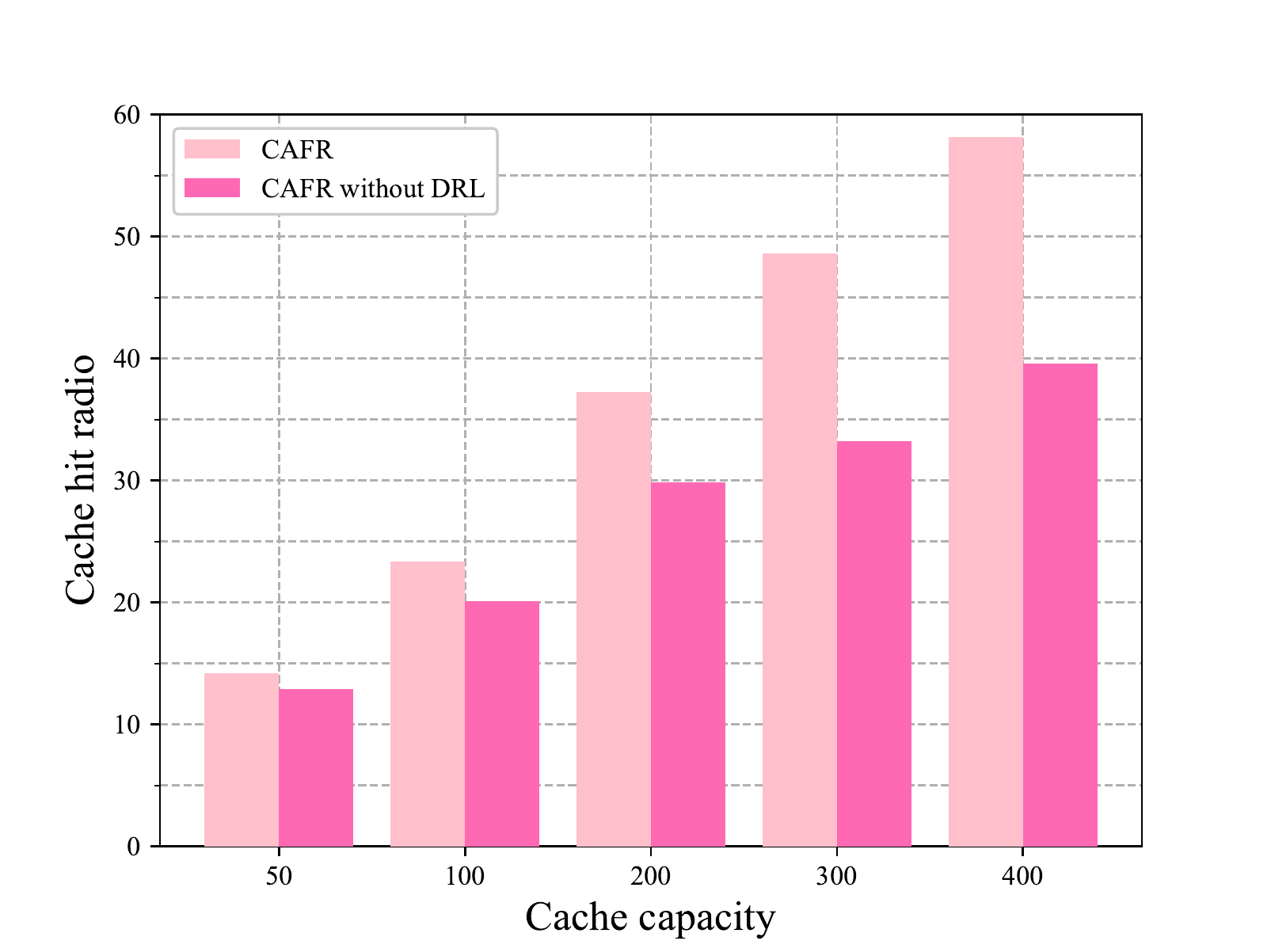}
\caption{Cache hit radio for whether cache replacement}
\label{fig11}
\end{figure}

Fig. \ref{fig11} compares the cache hit ratio of the CAFR scheme with CAFR scheme without DRL under different cache capacities of each RSU when the vehicle density is $15$ vehicles/km. As shown in Fig. \ref{fig11}, the cache hit ratio of CAFR outperforms the CAFR without DRL. This is because DRL can determine the optimal cooperative caching according to the predicted popular contents, and thus more suitable popular contents can be cached in the local RSU.

\begin{figure}
\center
\includegraphics[scale=0.5]{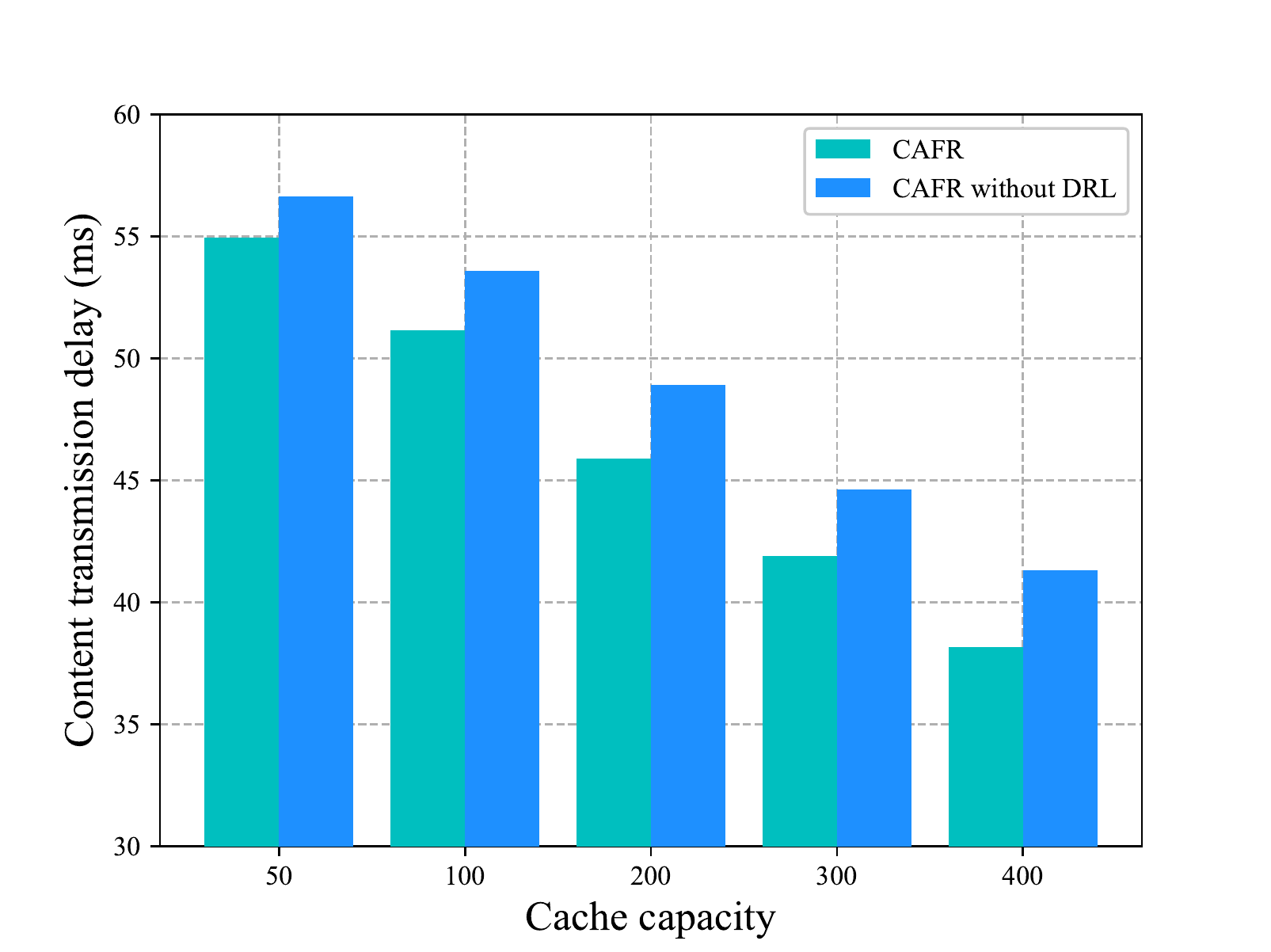}
\caption{Content transmission delay of CAFR and CAFR without DRL under different cache capacities}
\label{fig12}
\end{figure}

Fig. \ref{fig12} compares the content transmission delay of the CAFR scheme with CAFR scheme without DRL under different cache capacities of each RSU when the vehicle density is $15$ vehicles/km.
As shown in Fig. \ref{fig12}, the content transmission delay of CAFR is less than that of CAFR without DRL. This is because the cache hit ratio of CAFR outperforms the CAFR without DRL and more vehicles can fetch contents from local RSU directly.

%Combined with Fig. \ref{fig14} and \ref{fig15}, these experiment results show that the proposed cache replacement strategy based on dueling DQN can further improve the caching performance of the edge caching scheme in a highly dynamic vehicular environment.

\section{Conclusions}
\label{sec7}
In this paper, we considered the vehicle mobility and proposed a cooperative caching scheme CAFR to reduce the content transmission delay and improve the cache hit radio. We first proposed an asynchronous FL algorithm to obtain an accurate global model, and then proposed an algorithm to predict the popular contents based on the global model. Afterwards, we proposed a cooperative caching scheme to minimize the content transmission delay based on the dueling DQN algorithm. Simulation results have demonstrated that the CAFR scheme outperforms other baseline caching schemes. According to the theoretical analysis and simulation results, the conclusions can be summarized as follows:
\begin{itemize}
\item CAFR scheme can learn from the local data of vehicles to capture useful hidden features and predict the accurate popular contents.

\item CAFR greatly reduces the training time for each round by aggregating the local model of a single vehicle in each round. In addition, CAFR considers vehicles' mobility characteristics including the positions and velocities to select vehicles and aggregate the local model, which can improve the accuracy of the training model.

\item The DRL in the CAFR scheme determines the optimal cooperative caching policy according to the predicted popular contents, and thus more suitable popular contents are cached in the local RSU and neighboring RSU to reduce the content transmission delay.

\end{itemize}

%In this paper, we have proposed a mobility-aware cooperative edge caching scheme based on federated deep reinforcement learning (MCFR), to reduce the content transmission delay, improve cache hit radio and protect vehicles' privacy. To adapt to the highly dynamic VNs environment, MCFR adopts an asynchronous FL method to allow vehicles to train data locally. To predict content popularity, MCFR adopts Autoencoder model to train local data with hybrid filtering, which extracts potential correlations in the data from the vehicles' historical request contents and contextual information. MCFR utilizes dueling DQN model to dynamically update the cached contents of RSU based on the movement characteristics of mobility vehicles and the predicted content popularity. Numerical results show that MCFR outperforms other baseline caching schemes in terms of cache hit rate and average content transmission delay of each vehicle. The asynchronous FL training process of MCFR avoids stragglers and is better adapted to the highly dynamic VNs environment. The implementation of the cache placement strategy with dueling DQN further improves the cache hit rate and average content transmission delay of each vehicle.

\ifCLASSOPTIONcaptionsoff
  \newpage
\fi

%\begin{IEEEbiography}{Yuguang ``Michael'' Fang}
%Biography text here.
%\end{IEEEbiography}
%
%%It is not necessary to upload the biography when you submit your manuscript.
%
%
\end{document}

 %and able to control the buffer length near around the minimum expected buffer length.

% \section*{Acknowledgment}

% The authors would like to thank...

\ifCLASSOPTIONcaptionsoff
  \newpage
\fi

\end{document}